\documentclass[english,preprint,aps,superscriptaddress,showkeys,pre]{revtex4}
\usepackage[utf8x]{inputenc}
\usepackage{amsmath}
\usepackage{amssymb}
\usepackage{graphicx}
\usepackage{epstopdf}
\usepackage{setspace}

\makeatletter
\@ifundefined{textcolor}{}
{%
 \definecolor{BLACK}{gray}{0}
 \definecolor{WHITE}{gray}{1}
 \definecolor{RED}{rgb}{1,0,0}
 \definecolor{GREEN}{rgb}{0,1,0}
 \definecolor{BLUE}{rgb}{0,0,1}
 \definecolor{CYAN}{cmyk}{1,0,0,0}
 \definecolor{MAGENTA}{cmyk}{0,1,0,0}
 \definecolor{YELLOW}{cmyk}{0,0,1,0}
}

\usepackage[english]{babel}
\usepackage{subfigure}
 
\usepackage{amsfonts}\usepackage{textcomp}\usepackage{color}
\usepackage{array}

\makeatother

\usepackage{babel}
\begin{document}

\title{Chimera States in Hybrid Coupled Neuron Populations}

\author{Ali Calim}
\email[]{ali.calim@hotmail.com}
\affiliation{Department of Biomedical Engineering, Zonguldak Bulent Ecevit University, Zonguldak, Turkey}

\author{Joaquin J. Torres}
\affiliation{Department of Electromagnetism and Physics of the Matter and Institute Carlos I for Theoretical and Computational Physics, University of Granada, Granada, E-18071 Spain}

\author{Mahmut Ozer}
\affiliation{Ministry of National Education, Ankara, Turkey}
\affiliation{Center for Artificial Intelligence and Data Science, Istanbul Technical University, Istanbul, Turkey}
\affiliation{Department of Electrical and Electronics Engineering, Zonguldak Bulent Ecevit University, Zonguldak, Turkey}

\author{Muhammet Uzuntarla}
\affiliation{Department of Biomedical Engineering, Zonguldak Bulent Ecevit University, Zonguldak, Turkey}

\date{\today}

\begin{abstract}
Here we study the emergence of chimera states, a recently reported phenomenon referring to the coexistence of synchronized and unsynchronized dynamical units, in a population of Morris-Lecar neurons which are coupled by both electrical and chemical synapses, constituting a hybrid synaptic architecture, as in actual brain connectivity. This scheme consists of a nonlocal network where the nearest neighbor neurons are coupled by electrical synapses, while the synapses from more distant neurons are of the chemical type. We demonstrate that peculiar dynamical behaviors, including {chimera state and traveling wave}, exist in such a hybrid coupled neural system, and analyze how the relative abundance of chemical and electrical synapses affects the features of chimera and different synchrony states (i.e. incoherent, traveling wave and coherent) and the regions in the space of relevant parameters for their emergence. Additionally, we show that, when the relative population of chemical synapses increases further, a new intriguing chaotic dynamical behavior appears above the region for chimera states. This is characterized by the coexistence of two distinct synchronized states with different amplitude, and an unsynchronized state, that we denote as a chaotic amplitude chimera. We also discuss about the computational implications of such state.
\end{abstract}

%\pacs{05.45.Xt, 05.45.−a, 87.18.Sn}
\keywords{Chimera state, hybrid coupling, chaotic population behavior}

\maketitle

\section{Introduction}
Synchronization is widely considered to be essential for the proper functioning of a large variety of natural and artificial systems, ranging from physical experiments to chemical reactions and physiological processes. Prominent examples include communication networks \cite{nasir2016timing, hasan2018gnss}, coupled lasers \cite{kanno2017spontaneous, zhang2019cluster, heil2001chaos, wallace2000synchronization}, Josephson junctions \cite{galin2018synchronization, chitra2008phase}, oxidation and catalytic surface reactions \cite{lysak2016mathematical, nagiev2006coherent, salazar2004synchronization}, power grids \cite{nishikawa2015comparative} as well as circadian oscillators \cite{herzog2017regulating, buhr2013molecular} and genetic oscillator networks \cite{borg2014complex, wang2010synchronization, li2007stochastic}. Apart from these, synchronization in neural systems has remained a very popular research area during the last decades, because it is widely assumed to be a possible underlying mechanism for various behavioral and cognitive functions, e.g., attention, information processing, and neural control of movement \cite{sporns2000connectivity, womelsdorf2007role, velazquez2009coordinated, kelso2013outline, khanna2017beta}. Moreover, many findings from both experimental and theoretical research suggest that neural synchronization might be responsible for pathological conditions in brain diseases (i.e., epilepsy and Parkinson), where the synchronized oscillations are the significant difference between healthy and unhealthy conditions \cite{oswal2013synchronized, pollok2012motor, abd2015effects, li2019adjustment, hammond2007pathological, uhlhaas2006neural, batista2010delayed, uzuntarla2019synchronization}. Considering such important consequences, understanding the nature and controllability of neuronal synchronization is a critical step in uncovering the bases of many brain functions and diseases.

On the other hand, neural synchronization is not always desirable and ubiquitous in the brain \cite{hanslmayr2012oscillatory, waschke2019neural, nini1995neurons, magill2001dopamine}. It has been found that healthy brain exhibits spontaneous asynchronous activity as well as synchronous patterns \cite{ostojic2014two}. Thus, asynchronous population activity is not an harmful circumstance, it is rather beneficial to the brain. It helps for an efficient information processing and making decision in an excellent way, and also carrying out other vital tasks properly \cite{klimesch1997event, kitajima2013decision}. In particular, the cortex operates in a highly asynchronous state during waking and REM sleep \cite{steriade2013brainstem}. The subthalamic nucleus, a specific location in the basal ganglia, is another evidence of this inspection. It exhibits asynchronous electrical activity in the beta frequency band as an indicator of movement preparation \cite{heinrichs2013neuromagnetic}.

Recent experimental and clinical studies have shown that these two common states, namely, synchronous and asynchronous activity, can coexist within the same neuronal circuitry at the same time \cite{ahn2018neural, ahn2013dynamical}, and such a surprising state occurs, for instance, during unihemispheric sleep, epileptic seizures and bump states \cite{laing2001stationary, rattenborg2006birds, sakaguchi2006instability, truccolo2014neuronal, fard2015modeling, liou2018role}. In recent years, these evidences have motivated researchers from neurophysics community to study such coexisting states and relate them with physical phenomena observed in nonlinear dynamical systems. In this context, a widely considered representative dynamical phenomenon is the \textit{chimera state} which was originally described as coexistence of coherent and incoherent system states in a network of coupled identical phase oscillators with nonlocal interactions \cite{kuramoto2002coexistence, abrams2004chimera}. This symmetry-breaking physical concept has attracted great attention in determining the biological mechanisms that give rise to coexisting coherent and incoherent population activity in neural circuits. For instance, Omelchenko et al. have shown the emergence of chimera and multichimera -- {which refers to multiple incoherent domains} -- states in nonlocal network of electrically coupled Fitzhugh-Nagumo neurons \cite{omelchenko2013nonlocal}. To test robustness of their results, in \cite{omelchenko2015robustness}, authors further investigated chimera states in heterogeneous neuron population considering diversity of intrinsic excitability and coupling, and found that emergence of chimera states is robust for small heterogeneity but, as the heterogeneity increases, multichimeras transform into single chimera. In another work, Bera et al. explored chimera states in nonlocal, global, and local networks of chemically coupled bursting type Hindmarsh-Rose neurons \cite{bera2016chimera}, and found that chimera also occurs in population of such model neurons in the presence of chemical synapses at network interactions. In a recent work, we {have demonstrated} that populations of Morris-Lecar type model neurons also exhibit chimeric behavior with fine tuning of biophysically relevant parameters, i.e. excitability, synaptic strength and network connectivity \cite{calim2018chimera}. Apart from these works, presence of chimera state and its variants (e.g. amplitude chimera, breathing chimera and traveling chimera) {have been shown} in populations of other types of model neurons which are widely used in theoretical studies of neural circuits \cite{tsigkri2016multi, glaze2016chimera, sakaguchi2006instability}. These findings from modeling studies support the idea that emergence of chimera state can indeed be observable in actual neural circuits at the levels of cognitive and functional organizations \cite{bansal2019cognitive}.

In this work, we go a step further in the study and understanding of the appearance of chimera states in neural circuits by introducing  another biologically relevant condition, that is the existence of a hybrid synaptic architecture for interneuronal communication. {As is well-known from experimental findings, two main types of synapses, namely electrical and chemical ones, take part in synaptic transmission and neuron-to-neuron communication \cite{pereda2014electrical}}. At an electrical synapse, intercellular channels build a physical connection between cells, called gap junctions, and the signal transmission occurs through these channels directly from one neuron to another bidirectionally. However, information transfer across a chemical synapse take place unidirectionally from pre- to postsynaptic cell with complex biophysical mechanisms driving the dynamics of excitatory or inhibitory neurotransmitter particles released from presynaptic side which move across the synaptic cleft and activate receptor proteins on the postsynaptic neuron \cite{connors2004electrical}. There have been a large number of works revealing the presence of electrical synapses in different regions of the brain, such as the inferior olive \cite{llinas1974electrotonic}, locus coeruleus \cite{christie1989electrical}, hypothalamus \cite{ma2015electrical} and spinal cord \cite{chang1999gap}. On the other hand, chemical synapses are also common through the nervous system \cite{li2018dynamic}, and they are extensively found in different regions of cortex, hippocampus and olfactory bulb \cite{hormuzdi2004electrical, kennedy2016synaptic, dani2010superresolution}. Nevertheless, under the light of recent reports, it is now known that electrical and chemical synapses coexist in mammalian brain structures. Principal findings from neuroimaging and electrophysiological studies have showed that both forms of transmission can be simultaneously found at the same functional neural circuit, including retina \cite{kuo2016nonlinear}, neocortex \cite{smith2003chemical} and spinal cord \cite{rash1996mixed}.

So far, generic chimera studies concerning neuron populations have considered network connectivity formed with either solely electrical or chemical synapses. {Exceptionally, there only recently appeared a few studies investigating} effect of their coexistence on the emergence of chimeric behaviors in networks of networks. In such studies, neurons {communicate with} each other via one synapse type within a given network and via another type across different networks. For instance, Hizanidis et al. studied chimera states in modular neural networks and showed that chimera-like states spontaneously emerge with a suitable tuning of electrical and chemical coupling strengths within populations and across them, respectively \cite{hizanidis2016chimera}. On the other hand, Majhi et al. recently analyzed the chimera states in a two-layer neural network where connections between neurons are established via electrical synapses in one layer and chemical ones across the other target layer, and demonstrated that the emergence of chimera states depends significantly on coupling strengths of chemical synapses but poorly on the electrical ones \cite{majhi2017chimera}. However, these modeling approaches are not sufficient to address aforementioned biological reality, since hybrid synaptic connectivity is considered with lack of physiological findings. In the literature, to our knowledge, no attempts have been made to examine population behavior under consideration of hybridness associated with the connectivity in the same neural medium, except only one recent study carried out to assess the emergence of chimera state in a local community \cite{mishra2017traveling}. Although it is evaluated to be more reasonable to consider such a hybrid connectivity, this study has concentrated on only preliminary biophysical relevance, considering just locally electrical synapses as well as nonlocal chemical connections and Hindmarsh-Rose polynomial neuron model as in above mentioned previous works. Thus, it is worth looking at this subject from a wider perspective. In order to analyze in deep the role of coexisting chemical and electrical synapse populations for the emergence of chimera state, and to ensure more relevant and realistic assumptions, we here consider a modeling strategy for comprehensibility, such that chemical synapses are more common within the same neural circuitry and synapses from nearest neighbor neurons are of electrical type, whereas farther ones are of chemical type in a nonlocal network. Our main contribution in this work is to analyze emergence of chimera state in more physiological Morris-Lecar neuron populations coupled by abundant hybrid connections. We show that chemical synapses are essential for chimera-like behaviors whereas electrical ones are surprisingly a key component for emergence of new intriguing behavior in hybrid coupled network, namely chaotic amplitude chimera. 

The rest of the paper is organized as follows: In the next section, we introduce the neural population model, that is a set of $N=1000$ spiking Morris-Lecar neurons which are electrically and excitatory-chemically coupled in a nonlocal network, and the method used to characterize chimeric behavior, i.e. mean firing frequency. In the results section, we will first investigate how critical is the role that the relative number of each synapse type can play for the emergence of chimera state with given synaptic strengths. It is obvious that among different system features affecting the possible emergence of chimeric behaviors, synaptic coupling strength is one of the most significant factors in interneuronal communication since it dramatically affects the dynamics of the population. Consequently, as a next step, we will explore the influence of coupling strengths on the appearance of chimera-like states with a controlled variation for electrical and chemical synapses. After that, we also analyze the emergent intriguing chaotic behavior caused by the presence of hybrid synaptic interactions. Finally, our main findings and analysis are summarized in the conclusion section.

\section{Models and Methods}
\label{sec:model}
\begin{figure}[t]
\begin{centering}
\includegraphics[trim={1.5cm 1.0cm 1.5cm 0.90cm},clip,height=3.95cm]{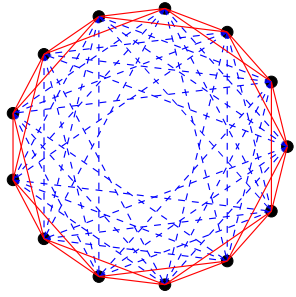}
\end{centering}
\caption{Figure shows the scheme of nonlocal hybrid connectivity used in the present study. In the plotted example there are $13$ neurons in the network in the form of a ring where each one is connected to $R=2$ neighbors via electrical synapses (red solid lines) and to $S=3$ neighbors via chemical connections (blue dashed lines).}
\label{fig1}
\end{figure}

Lets consider a network of coupled neurons placed in the nodes of a ring as it is depicted in Fig. \ref{fig1}. The dynamics of the membrane potential of each neuron in the network is modeled using the two-variable Morris-Lecar equations \cite{morris1981voltage, Rinzel1989ANE, uzuntarla2013inverse, shein2016modularity}:
\begin{eqnarray}
\label{eq:ml_model}
C \frac{dV_i}{dt} &=& \,I_0 + g_{\text{Ca}}m_i^{\infty}(E_{\text{Ca}}-V_i) + g_{\text{K}}w_i(E_{\text{K}}-V_i) + g_{\text{L}}(E_{\text{L}}-V_i) + I_i^{\text{syn}}\\
\frac{dw_i}{dt} &=& \,\phi (w_i^{\infty}-w_i) \, \cosh\left(\frac{V_i-\beta_w}{2\gamma_w}\right)\\
m_i^{\infty}(V_i) &=& \,0.5\left[1+\tanh\left(\frac{V_i-\beta_m}{\gamma_m}\right)\right] \\
w_i^{\infty}(V_i) &=& \,0.5\left[1+\tanh\left(\frac{V_i-\beta_w}{\gamma_w}\right)\right],
\end{eqnarray}
where $i=1,2,\dots,N$ denotes the neuron index. {$V_i$ and $w_i$ represent the membrane potential and activation dynamics of potassium channels for neuron $i$, respectively. $I_0$ is a constant bias current externally applied to all neurons in the network, which is fixed to $I_0=10\,\mu A/cm^2$ providing regularly spiking individual cells in the population. The parameters $w_i^\infty$ and $m_i^\infty$ are the steady-state functions of activated potassium and calcium channels, respectively.} The constants $g_{\text{Ca}}=1\,$\textit{mS/cm}$^2$, $g_{\text{K}}=2\,$\textit{mS/cm}$^2$ and $g_{\text{L}}=0.5\,$\textit{mS/cm}$^2$ are maximal conductance values for calcium, potassium and leak channels, respectively. Accordingly, $E_{\text{Ca}}=100\,$\textit{mV}, $E_{\text{K}}=-70\,$\textit{mV} and $E_{\text{L}}=-50\,$\textit{mV}  represent the corresponding ionic equilibrium potentials. Other system parameters are set as $C=1\,\mu$\textit{F/cm}$^2$ (the cell membrane capacitance), $\phi=1/3$, $\beta_m=-1\,$\textit{mV}, $\gamma_m=15\,$\textit{mV}, $\beta_w=10\,$\textit{mV} and $\gamma_w=14.5\,$\textit{mV}.

In Eq.~(1), $I_i^{\text{syn}}$ denotes the total synaptic current received by neuron $i$ from its neighbors in the ring. In order to connect neurons, we consider here a hybrid coupling scheme with electrical and chemical synapses incorporated into a nonlocal network as shown in Fig.~\ref{fig1}. More precisely, we consider that each neuron in such networked ring is electrically connected with its $2R$ nearest neighbors neurons in the ring and excitatory chemically coupled with $2S$ more distant neurons. This strategy results in totally $2(R+S)$ connections for each neuron coupled electrically to $R$ and chemically to $S$ neighbors in both directions as illustrated in Fig.~\ref{fig1}. Then, the total synaptic current a neuron is receiving from its neighbors can be written as $I_i^{\text{syn}}=I_i^E+ I_i^C$ with
\begin{equation}
I_i^{E}=\frac{1}{2R}\sum_{j=i-R}^{j=i+R} g_e  (V_j-V_i)
\end{equation} 
\begin{equation}
I_i^{C}=\sum_{j=i-R-S,\, j \neq i-R}^{j=i+R+S,\, j \neq i+R} g_c \, y_j
\end{equation}
where $g_e$ is the electrical coupling strength and $g_c$ is the maximum postsynaptic current which can be generated at the synapse by activating all synaptic resources. When a spike arrives at a chemical synapse $j$ at time $t$, there is an instantaneous release of a fraction $u_j=0.9$ of neurotransmitter resources that then becomes active to transmit the spike. Active resources, namely $y_j(t),$ then deactivate over a time on the order of a few milliseconds, characterized by the time constant $\tau_{in}$. We fixed it as $\tau_{in}=10\,ms$ for whole subsequent study, which is within the physiological range for excitatory synapses \cite{kaiser2007molecular, o2017beyond, avermann2012microcircuits}. Using standard synaptic transmission modeling \cite{panzeri2001speed, inbook}, we assume that the dynamical behavior of the fraction of active neurotransmitter resources $y_j(t)$ is governed by the following dynamics:
\begin{equation}
\frac{dy_j}{dt} = - \frac{y_j}{\tau_{in}} + u_j\delta(t-t_j^{AP})
\end{equation}
where the delta function refers to the arrival time of a spike at synapse $j$ at $t=t_j^{AP}$, which is defined by the upward crossing of the membrane potential past a threshold of $10\,mV$. 

\begin{figure}[!t]
\begin{minipage}{0.23\textwidth}
\includegraphics[scale=0.7521,trim={1mm 0 00mm 0},clip=true]{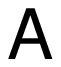}\\
\vspace{00.11cm}
\includegraphics[scale=0.521]{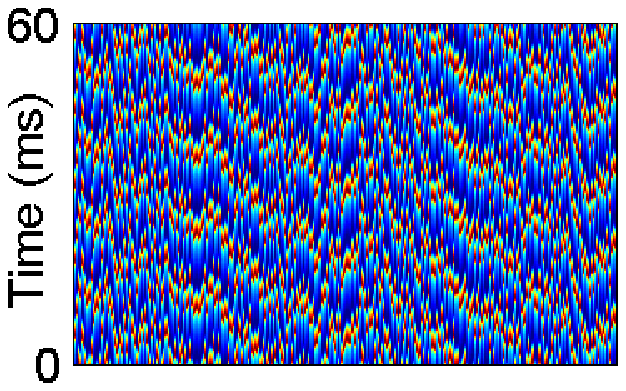}\\
\vspace{-00.40cm}
\includegraphics[scale=0.521]{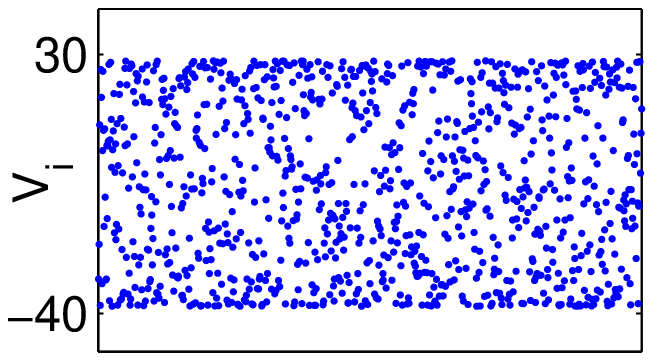}\\
\vspace{-00.40cm}
\includegraphics[scale=0.521]{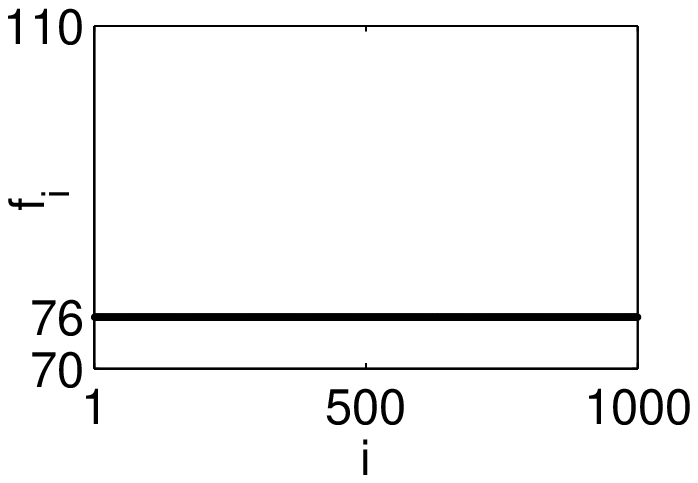}\\
\includegraphics[scale=0.521]{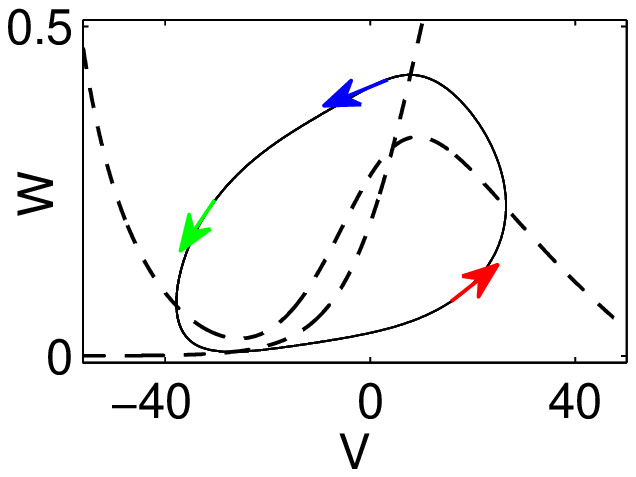}
\end{minipage}
\begin{minipage}{0.23\textwidth}
\includegraphics[scale=0.7521,trim={1mm 0 00mm 0},clip=true]{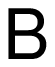}\\
\vspace{00.11cm}
\includegraphics[scale=0.521]{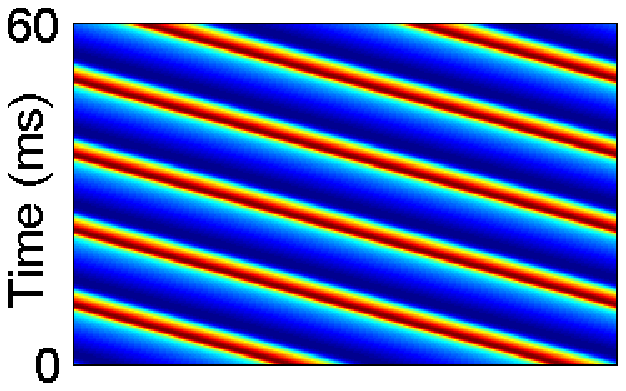}\\
\vspace{-00.40cm}
\includegraphics[scale=0.521]{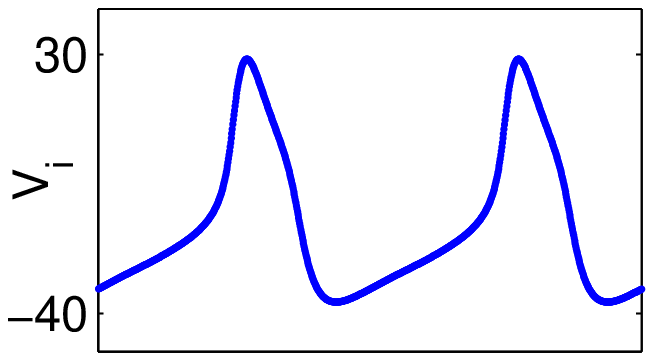}\\
\vspace{-00.40cm}
\includegraphics[scale=0.521]{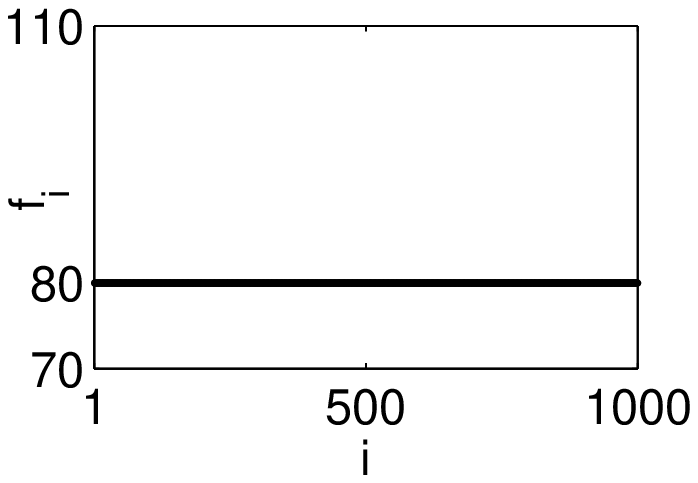}\\
\includegraphics[scale=0.521]{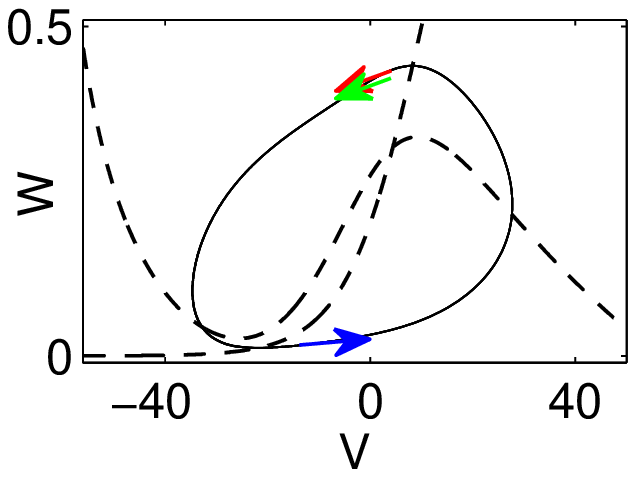}\\
\end{minipage}
\begin{minipage}{0.23\textwidth}
\includegraphics[scale=0.7521,trim={1mm 0 00mm 0},clip=true]{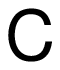}\\
\vspace{00.11cm}
\includegraphics[scale=0.521]{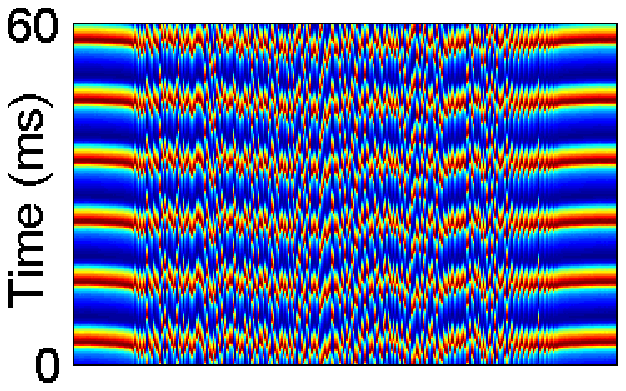}\\
\vspace{-00.40cm}
\includegraphics[scale=0.521]{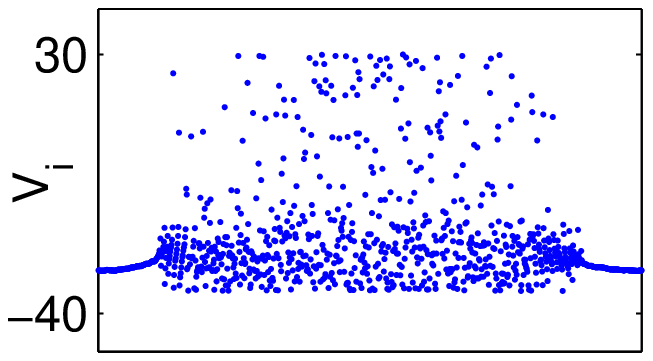}\\
\vspace{-00.40cm}
\includegraphics[scale=0.521]{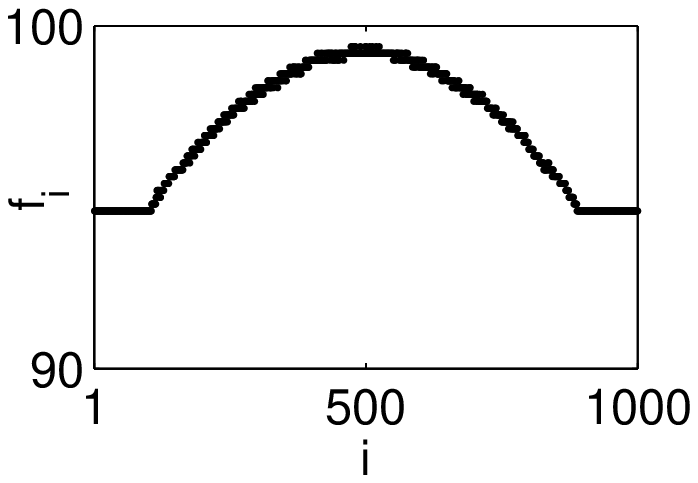}\\
\includegraphics[scale=0.521]{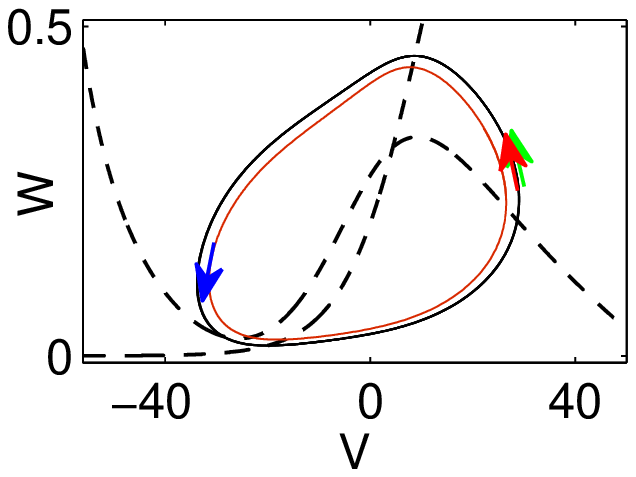}\\
\end{minipage}
\begin{minipage}{0.23\textwidth}
\includegraphics[scale=0.7521,trim={1mm 0 00mm 0},clip=true]{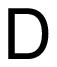}\\
\vspace{00.11cm}
\includegraphics[scale=0.521]{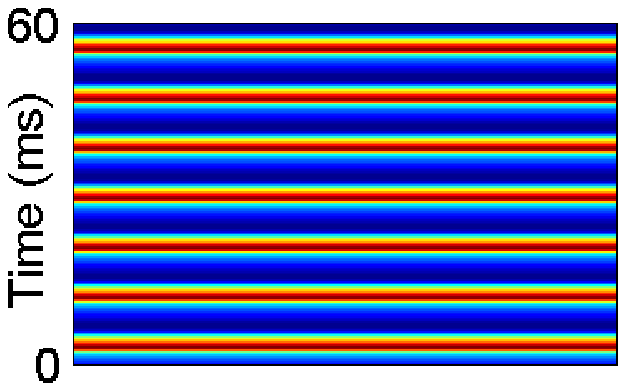}\\
\vspace{-00.40cm}
\includegraphics[scale=0.521]{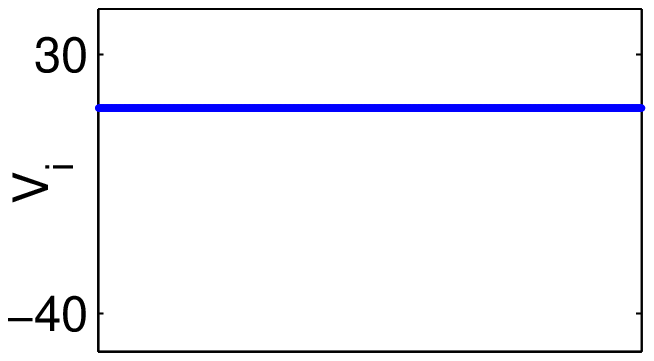}\\
\vspace{-00.40cm}
\includegraphics[scale=0.521]{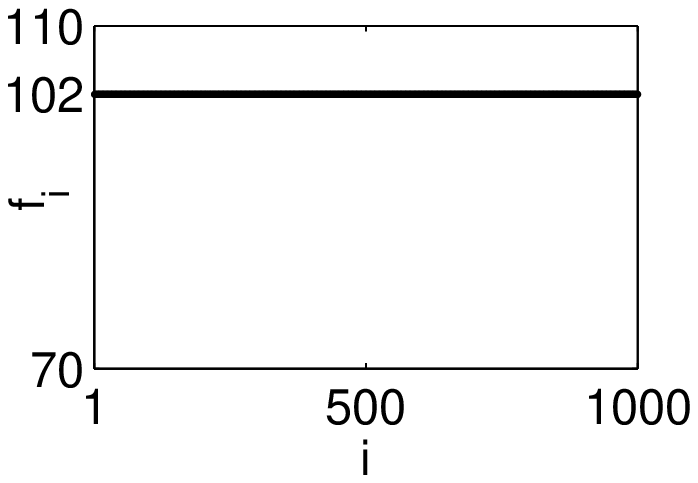}\\
\includegraphics[scale=0.521]{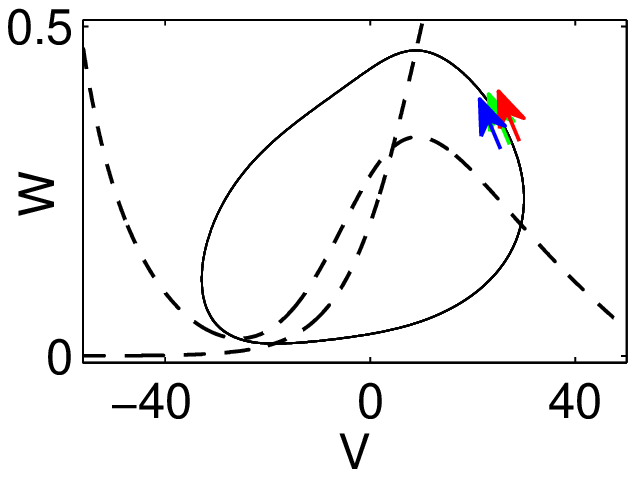}\\
\end{minipage}
\begin{minipage}{0.04\textwidth}
\includegraphics[scale=0.521,trim={16mm 0 6mm 0},clip=true]{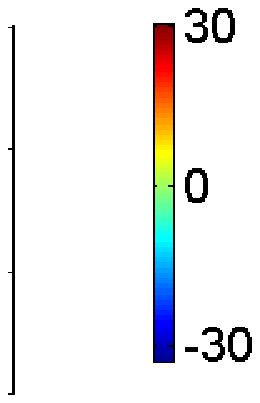}\\
\vspace{6.40cm}
\end{minipage}
\caption{Emergence of different dynamical behaviors in a hybrid coupled neural population as described in Fig. \ref{fig1} with variation of the number $S$ of chemical connections with a fixed number of electrical connections $R$. Each row shows spatiotemporal activity patterns, snapshots of membrane potentials, mean firing frequency profiles, and periodic orbits with instantaneous positions of two neighboring neurons in the networked ring (marked with red and green arrows) and one distant neuron  (marked with blue arrow) on $V-w$ phase plane, respectively. Number of chemical connections are set as $S=5$ (A), $S=125$ (B), $S=250$ (C) and $S=350$ (D). Other system parameters are fixed as $g_c=10^{-2}\,$\textit{mS/cm}$^2$, $g_e=10^{-7}\,$\textit{mS/cm}$^2$ and $R=100$.}
\label{fig2}
\end{figure}

To quantitatively determine the population activity behavior and characterize the existence of chimera states, in the following we will monitor the behavior of the mean firing frequency of all neurons in the ring, which is defined as $f_i = F_i/\Delta T$ for any given parameter set. Here, $F_i$ is the number of spikes fired by neuron $i$ within a period of time $\Delta T$ computed after a sufficient transient time. {The initial conditions for Eqs.~(1-7) are randomly selected with uniform probability within fixed intervals of $(-40\,\textit{mV}, 30\,\textit{mV})$ for $V_i$, $(0, 0.4)$ for $w_i$ and $(0, 1)$ for $y_i$.} Numerical integration of our system is performed using the fourth-order Runge-Kutta algorithm with a fixed time step of $10\, \mu$\textit{s}.

\section{Results}

In the following, we systematically investigate the emergent dynamical behaviors, especially chimera-like states, in hybrid coupled spiking neural populations as described in the previous section. As a first step, we begin by demonstrating the appearance of several distinct population behaviors when the chemical connection density $S$ is varied for a particular fixed number of electrical connections (that we set to $R=100$) with maximal conductances for electrical and chemical synaptic current being, respectively, $g_e=10^{-7}\,$\textit{mS/cm}$^2$ and $g_c=10^{-2}\,$\textit{mS/cm}$^2$. The corresponding obtained results are illustrated in  Fig.~\ref{fig2} where panels in each column, from top to bottom, show spatiotemporal patterns, snapshots of membrane potentials, mean firing frequencies, and periodic orbits with instantaneous positions (marked with colored arrows) of selected three neurons projected on $V$-$w$ phase plane, respectively. 

{By visual inspection of the spatiotemporal patterns and membrane potential snapshots, it is obvious that hybrid coupled population exhibits four different dynamical behaviors as $S$ increases. First one is the incoherent state where neurons fire independently. Each neuron evolves with regard to its initial position in parameter space without waiting any response from neighboring neurons. Second behavior is the intriguing activity pattern of traveling wave, which consists of spatially coherent oscillations that propagate progressively across the population. This type of activity widely occurs in different oscillatory brain states and under different sensory conditions, and is associated with  transmission of neural information across different functional brain regions, for example, during propagation of theta and alpha band rhythms \cite{zhang2018theta} and spread of epileptic seizures \cite{bertram2013neuronal}. Next, third one is the intriguing population behavior of chimera state. This state describes the occurrence of synchronous and asynchronous electrical activity in the same functional healthy or diseased brain regions  \cite{kang2019two}. Finally, the fourth population behavior corresponds to a coherent state where neurons fire in a synchronous and phase-locked manner. This last behavior is widely assumed to be a critical mechanism for various vital functions of nervous system, such as information processing and transmission \cite{varela2001brainweb, abeles1994synchronization}, movement control \cite{mamun2015movement} and many other different cognitive or behavioral tasks \cite{uhlhaas2008role}.}

To quantitatively characterize these different behaviors, we compute mean firing frequencies of individual neurons in the population as shown in third panels of each column in Fig. \ref{fig2}. We observe that all neurons, for a given population state, fire at a constant frequency, except for chimera state which has a characteristic bell-shaped mean firing frequency profile indicating the coexistence of two different subpopulations, coherent and incoherent, within the same network. It is also worth to note that mean firing frequency of the hybrid coupled population increases with $S$ {regardless of the existing dynamical state in which the system operate.}

For a more clear understanding of the above-mentioned emergent behaviors, we also perform a phase plane analysis (for each one of the illustrated cases) of the activity trajectories of three particular neurons from the population, which are selected as two neighbors $i=1,2$ in the networked ring and a distant neuron  $i=200$. This is depicted in the bottom panels of each column of Fig. \ref{fig2} where it is seen that these three neurons move on a single orbit in incoherent, traveling wave, chimera state and coherent states, respectively when $S$ is increased. One can easily distinguish these states by following the trajectories of each cell (marked with arrows) in $V$-$w$ phase plane. Although the phase plane behavior of three neurons in the traveling wave and chimera state seems to be similar, we observe that, in the chimera state, the phase profile of neighboring neurons from coherent group differs from that of the distant one belonging to the incoherent group, in such a way that coherent and incoherent group trajectories move on two different periodic orbits. We see more distinct periodic orbits in the phase plane when additional different neurons from incoherent subpopulation are considered (not shown for simplicity). This is a clear indicator for the presence of coherent and incoherent subgroups having different mean firing rates within the same population.

\begin{figure}[t]
\begin{minipage}{0.325\textwidth}
\includegraphics[scale=0.521,trim={4mm 0 0mm 0},clip=true]{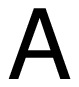}\\
\vspace{3mm}
\hspace{-0.25cm}\includegraphics[scale=0.37]{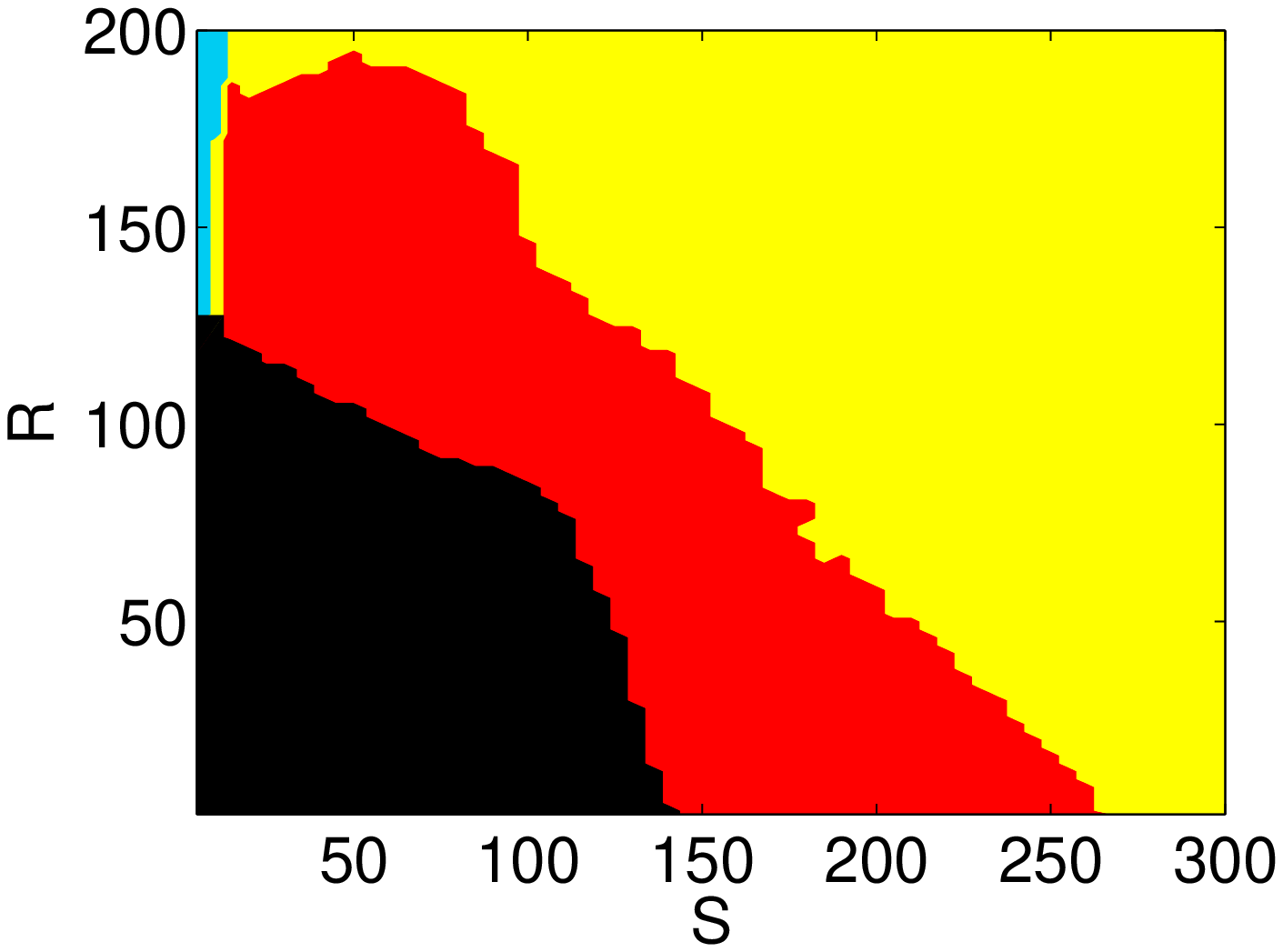}\\
\end{minipage}
\begin{minipage}{0.325\textwidth}
\includegraphics[scale=0.521,trim={4mm 0 0mm 0},clip=true]{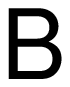}\\
\vspace{3mm}
\hspace{-0.25cm}\includegraphics[scale=0.37]{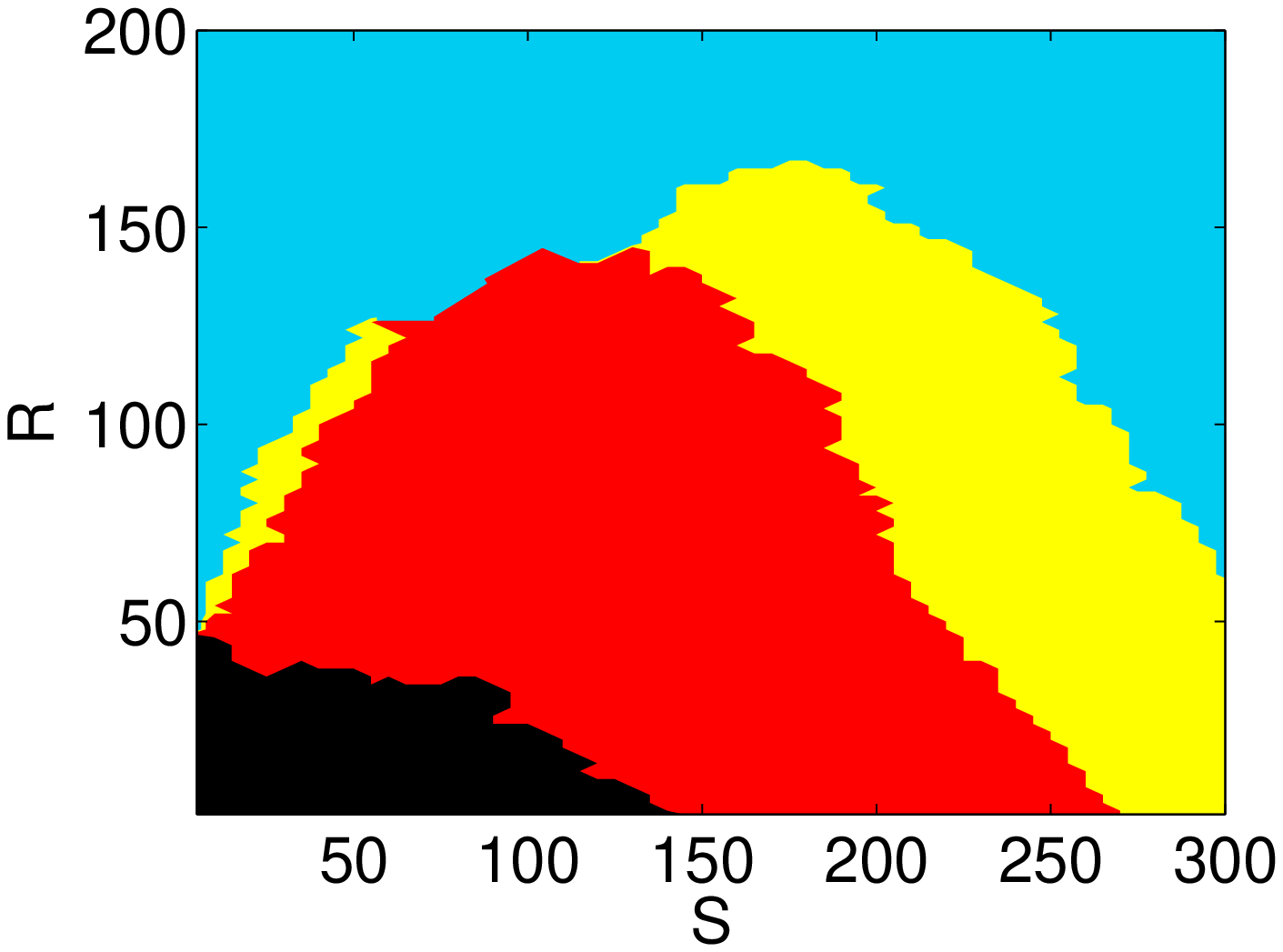}\\
\end{minipage}
\begin{minipage}{0.325\textwidth}
\includegraphics[scale=0.521,trim={4mm 0 0mm 0},clip=true]{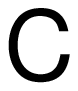}\\
\vspace{3mm}
\hspace{-0.25cm}\includegraphics[scale=0.37]{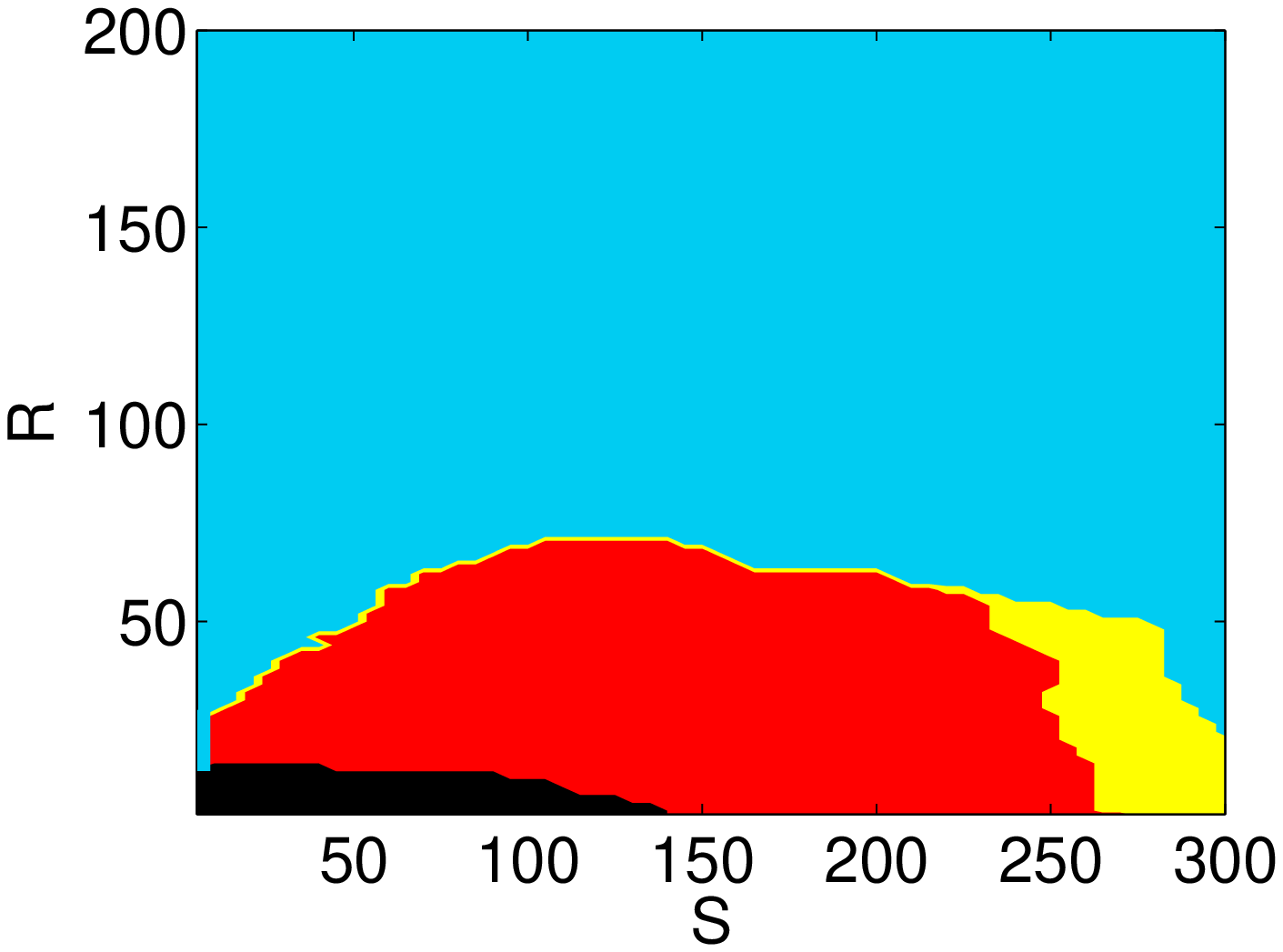}\\
\end{minipage}
\caption{Each panel shows phase diagrams spanned by parameters $R$ and $S$ for different electrical coupling strengths, fixed as $g_e=10^{-7}\,$\textit{mS/cm}$^2$ (A), $g_e=10^{-6}\,$\textit{mS/cm}$^2$ (B) and $g_e=10^{-5}\,$\textit{mS/cm}$^2$ (C). Regions for different dynamic behavior are given by the following color codes, Black: Incoherent state, Red: Traveling wave, Yellow: Chimera state and Blue: Coherent state. Chemical coupling strength is set to $g_c=10^{-2}\,$\textit{mS/cm}$^2$.}
\label{fig3}
\end{figure}

The above results clearly demonstrate that just tuning the single system parameter $S$ (the relative abundance of chemical synapses) can induce the appearance of non-trivial dynamical behaviors in the system, i.e. traveling wave and chimera states. In the following, we analyze in deep how such dynamical states can emerge in the considered hybrid coupling scheme as a function of other system parameters. Firstly, we analyze hybrid coupled population behavior on ($R$, $S$) plane for three different $g_e$ values as depicted in Fig.~\ref{fig3}, which provides us a broader perspective and confirmation of robustness for these observed intriguing dynamical states. When $g_e$ is small (see Fig.~\ref{fig3}A), we observe all above-mentioned population behavior types on ($R$, $S$) plane with large regions for incoherent, traveling wave and chimera states, and with a very narrow region for coherent state. This is mainly due to the small effect of the electrical synaptic current to induce synchronization of neuron activities in the population since we have $g_e\ll 1\,$\textit{mS/cm}$^2$. However, even in this case,  the situation is far to be $g_e$-independent since the transition lines between different types of dynamical behavior show a non-trivial inverse relationship between $R$ and $S,$ which is a clear mark of the presence of some electrical synapse mediated current effect.

For larger values of $g_e,$ one can also see that these regions are significantly modulated (see Fig.~\ref{fig3}B and C) where incoherent and chimera state regions get smaller while the coherent state region enlarges dramatically in the ($R$, $S$) space as $g_e$ increases. However, the overall shape of the region for the emergence of traveling waves does not change very much although it shifts towards lower $R$ values. At low $g_e$, it is obvious that both connection densities $R$ and $S$ are jointly responsible for the emergence of traveling wave and chimera states. But increasing electrical coupling strength apparently disrupts this balance and behavioral variety can be obtained with only very few electrical connections (low $R$) depending on the number of chemical synapses $S$. On the other hand, these results reveal that electrical synapses in hybrid coupled population are in favor of establishing coherent and incoherent states while chemical synapses promote all emergent behaviors except coherent state. This can be inferred by following behavioral maps on $R$-axes ($S$-axes) for $S=0$ ($R=0$) as depicted in Fig. \ref{fig3}A, B and C.

\begin{figure}[t]
\includegraphics[scale=0.41,trim={00mm 0 4mm 0},clip=true]{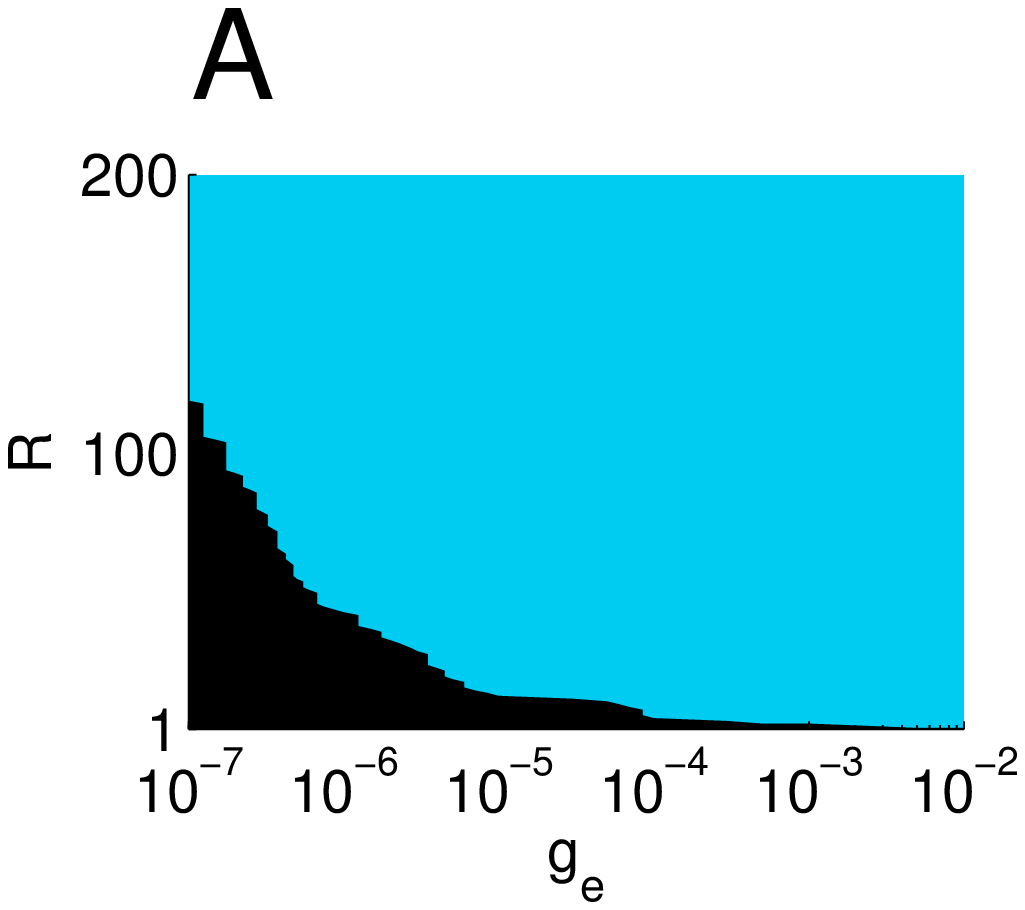}\hspace{00.11cm}
\includegraphics[scale=0.41,trim={10mm 0 4mm 0},clip=true]{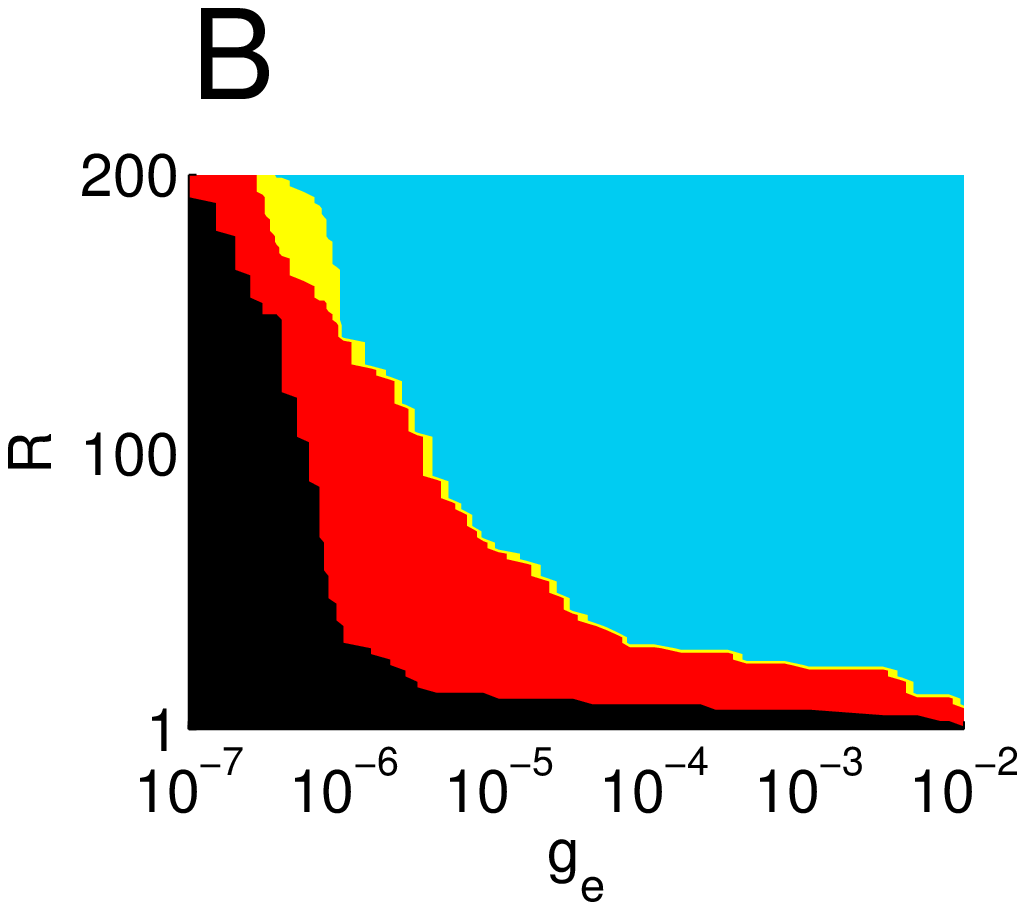}\hspace{00.11cm}
\includegraphics[scale=0.41,trim={10mm 0 4mm 0},clip=true]{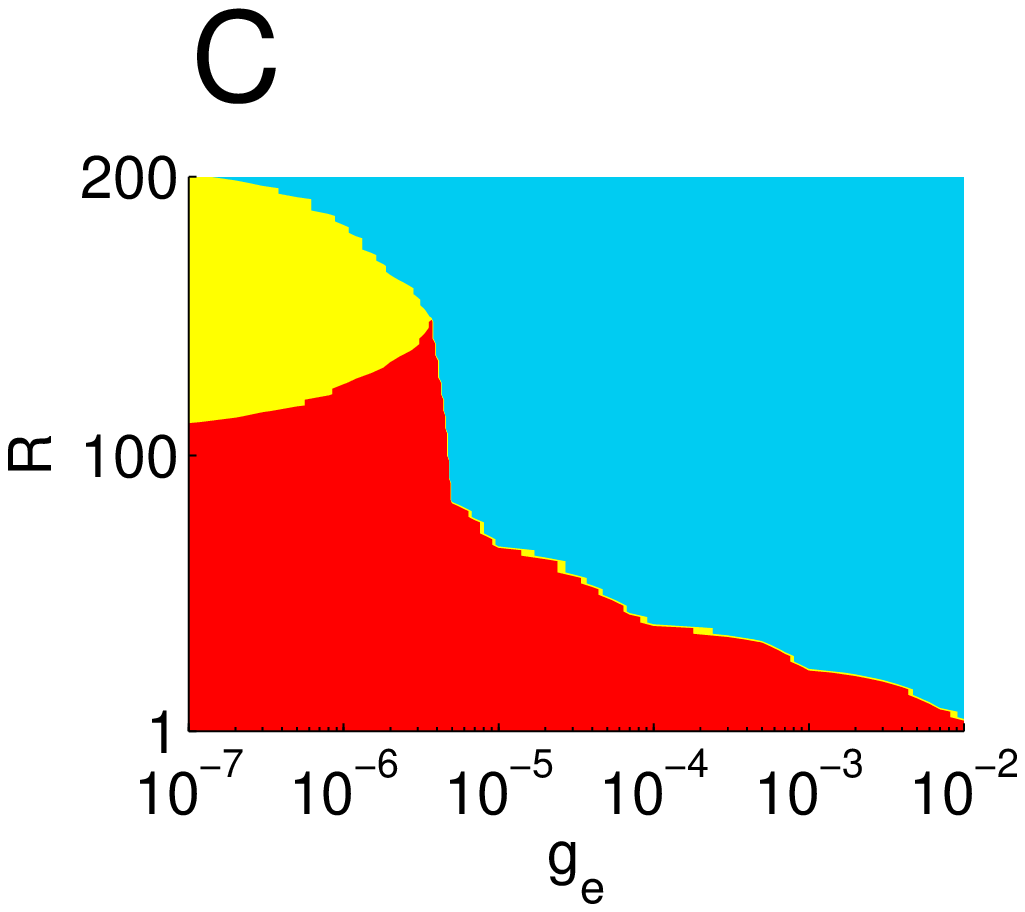}\\
\includegraphics[scale=0.41,trim={00mm 0 4mm 0},clip=true]{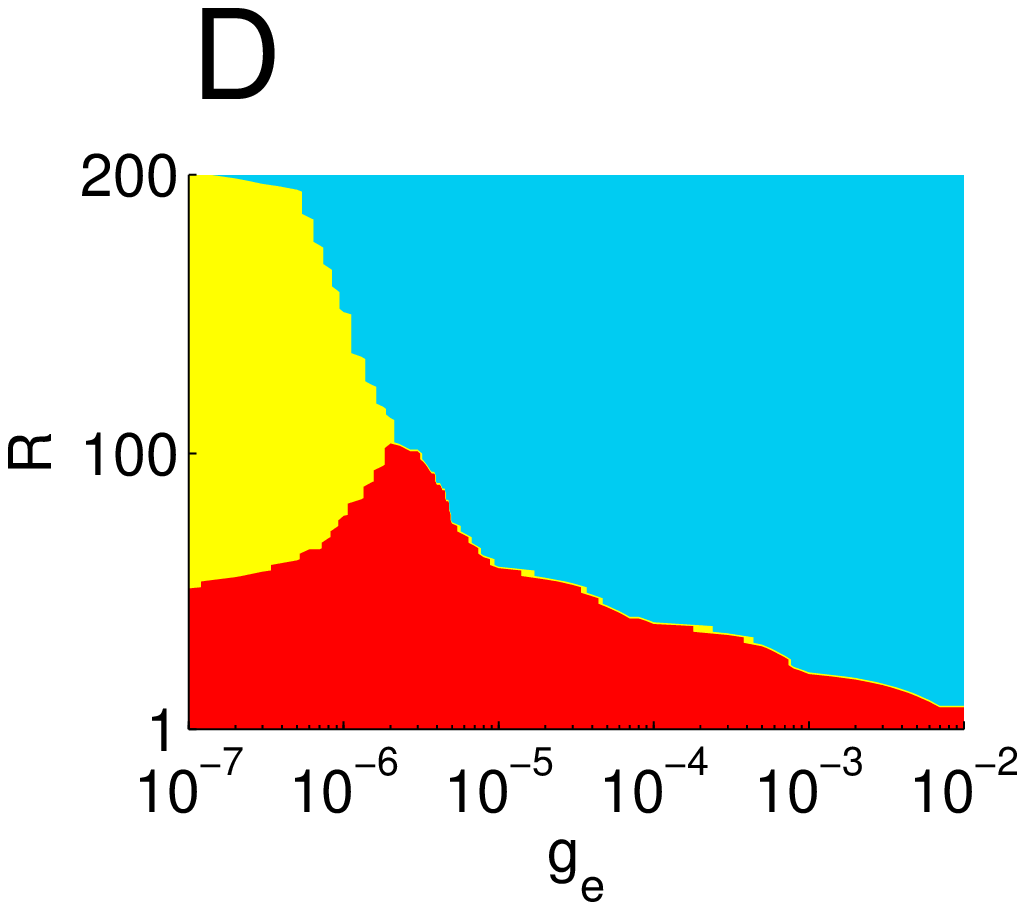}\hspace{00.11cm}
\includegraphics[scale=0.41,trim={10mm 0 4mm 0},clip=true]{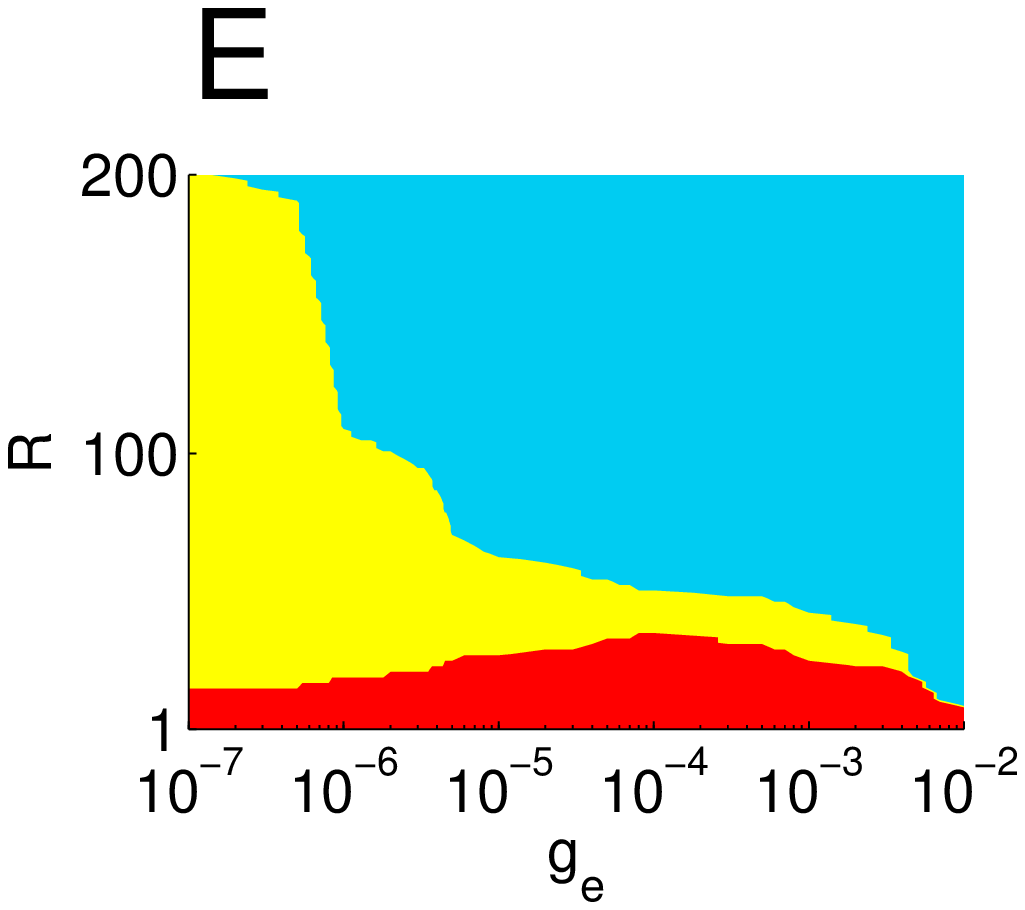}\hspace{00.11cm}
\includegraphics[scale=0.41,trim={10mm 0 4mm 0},clip=true]{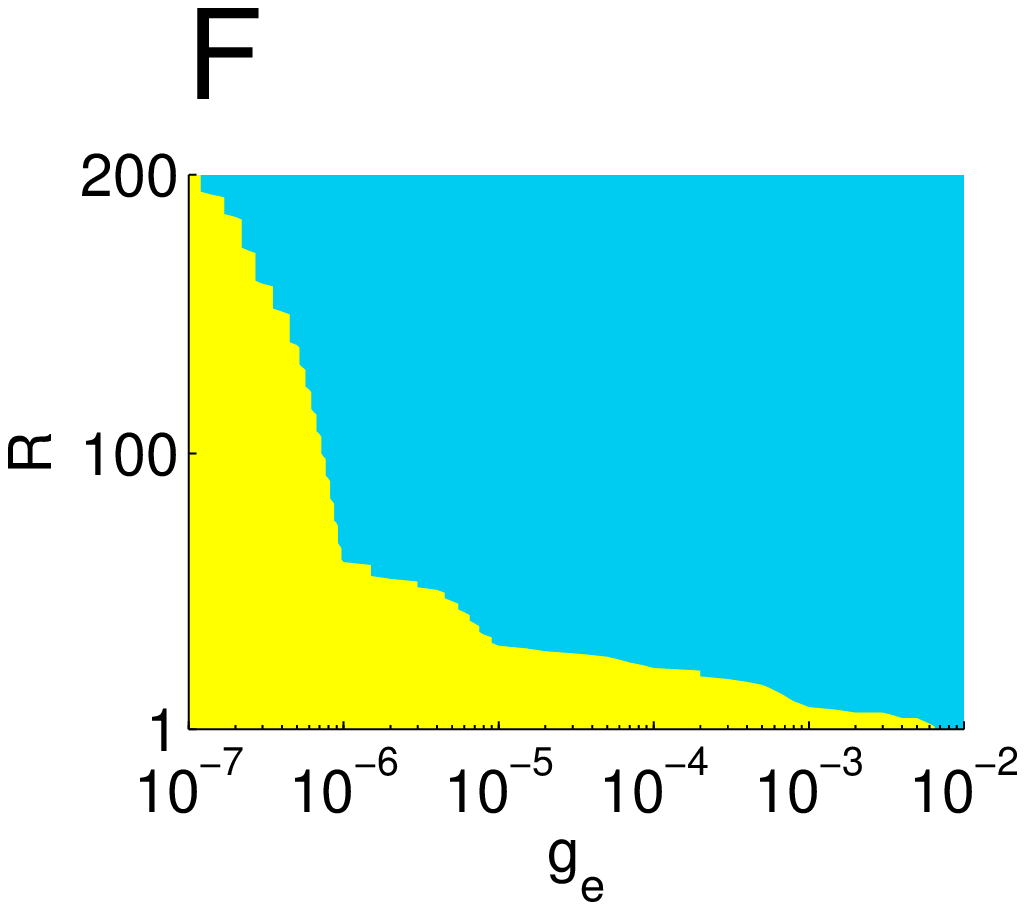}\\
\caption{Emergent dynamical behaviors in a hybrid coupled population as a function of electrical synapse features ($R$ and $g_e$) for different levels of chemical connection density $S$. Considering a fixed coupling strength for chemical synapses ($g_c=10^{-2}\,$\textit{mS/cm}$^2$), analysis have been performed for $S=5$ (A), $S=100$ (B), $S=150$ (C), $S=200$ (D), $S=250$ (E) and $S=300$ (F). Note that each population state is represented with different color codes being Black: Incoherent state, Red: Traveling wave, Yellow: Chimera state and Blue: Coherent state.}
\label{fig4}
\end{figure}

To gain more insight into the dependence of hybrid coupled population behavior on electrical synapses, we jointly scan a wide interval for electrical coupling strength $g_e$ and connection number $R$ in parameter space for different chemical connection densities $S$. Obtained results are illustrated in Fig.~\ref{fig4}. In the presence of very small number of chemical synapses (see Fig.~\ref{fig4}A for $S=5$), we observe only incoherent and coherent population behaviors for lower and higher electrical interaction intensities, respectively. Note that here, \textit{interaction intensity} refers to the combined effect of both $R$ and $g_e$. Figure~\ref{fig4}A also generalizes the behavior depicted on Fig. \ref{fig3} for low $S$ values as a function of $R,$ implying that electrical synapses are not so important in producing traveling wave and chimera state when $S$ is very low. However, increasing the number of chemical synapses in the population reveals rich behavioral variety on ($R$, $g_e$) parameter space. For instance, in addition to incoherent and coherent states, traveling wave and chimera state start to appear in population behavior when $S$ is increased to $100$ (see Fig.~\ref{fig4}B). Interestingly, a further increase of $S$ results in disappearance of incoherent state region, and chimera state and traveling wave extend over that region of the parameter space (Fig.~\ref{fig4}C). Then, chimera state starts to be the predominant dynamical behavior while traveling wave fades away with further increase of $S$ (see Fig. \ref{fig4}D, E and F for increasing values of $S$). It is worth to note that, despite the influence of $S$ in modulating the regions for incoherent, traveling wave and chimera states in parameter space, the region for occurrence of coherent state does not change very much with increased number of chemical synapses.

\begin{figure}[t]
\includegraphics[scale=0.410,trim={00mm 0 5mm 0},clip=true]{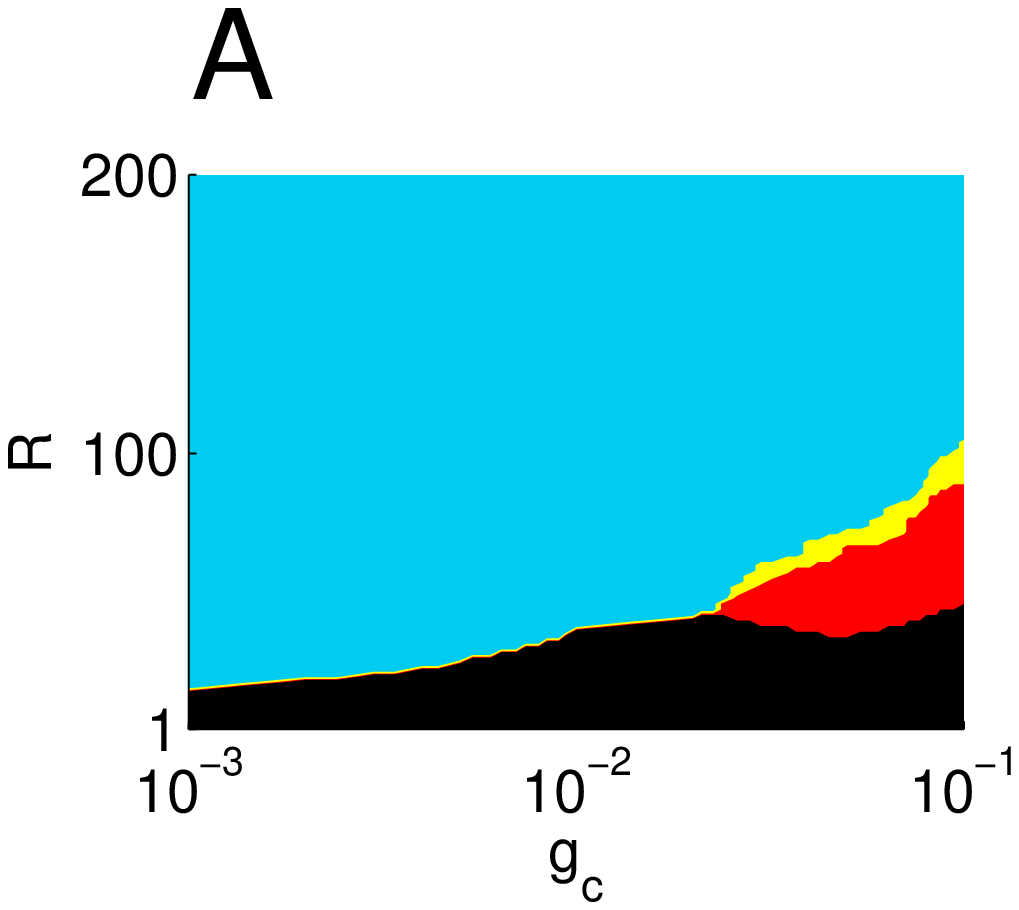}\hspace{00.025cm}
\includegraphics[scale=0.410,trim={10mm 0 5mm 0},clip=true]{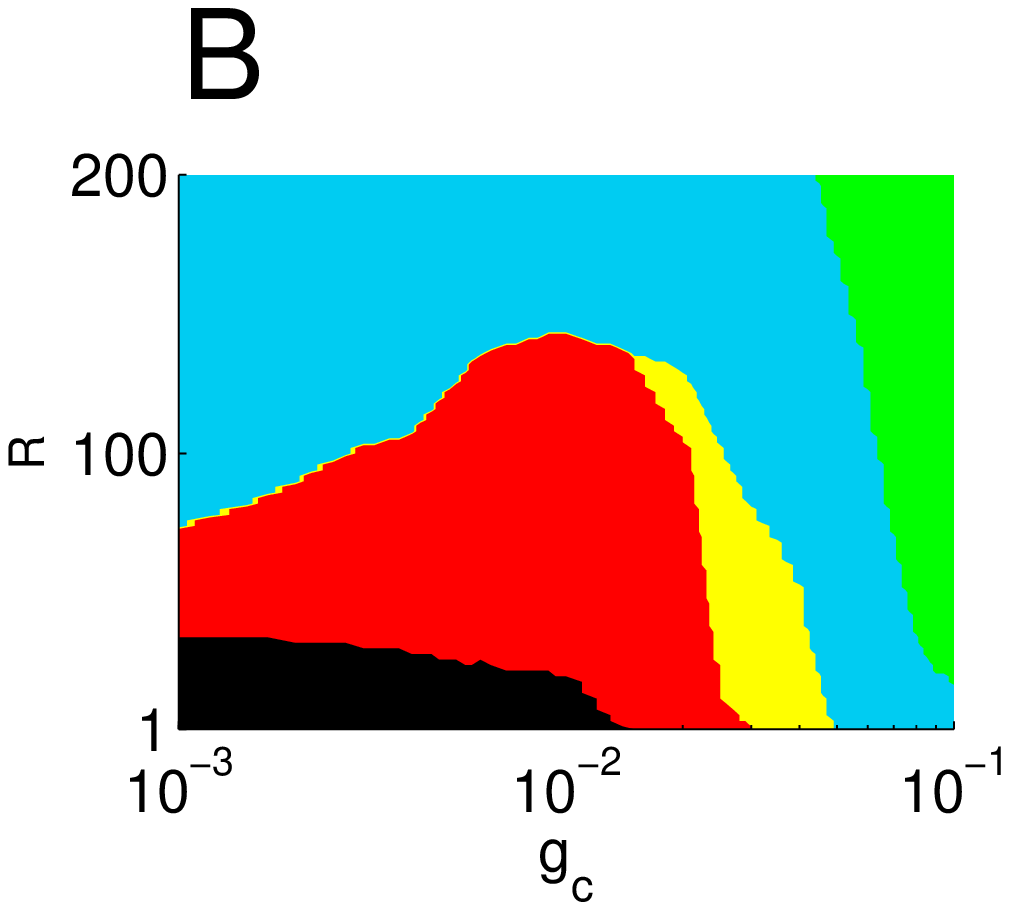}\hspace{00.025cm}
\includegraphics[scale=0.410,trim={10mm 0 5mm 0},clip=true]{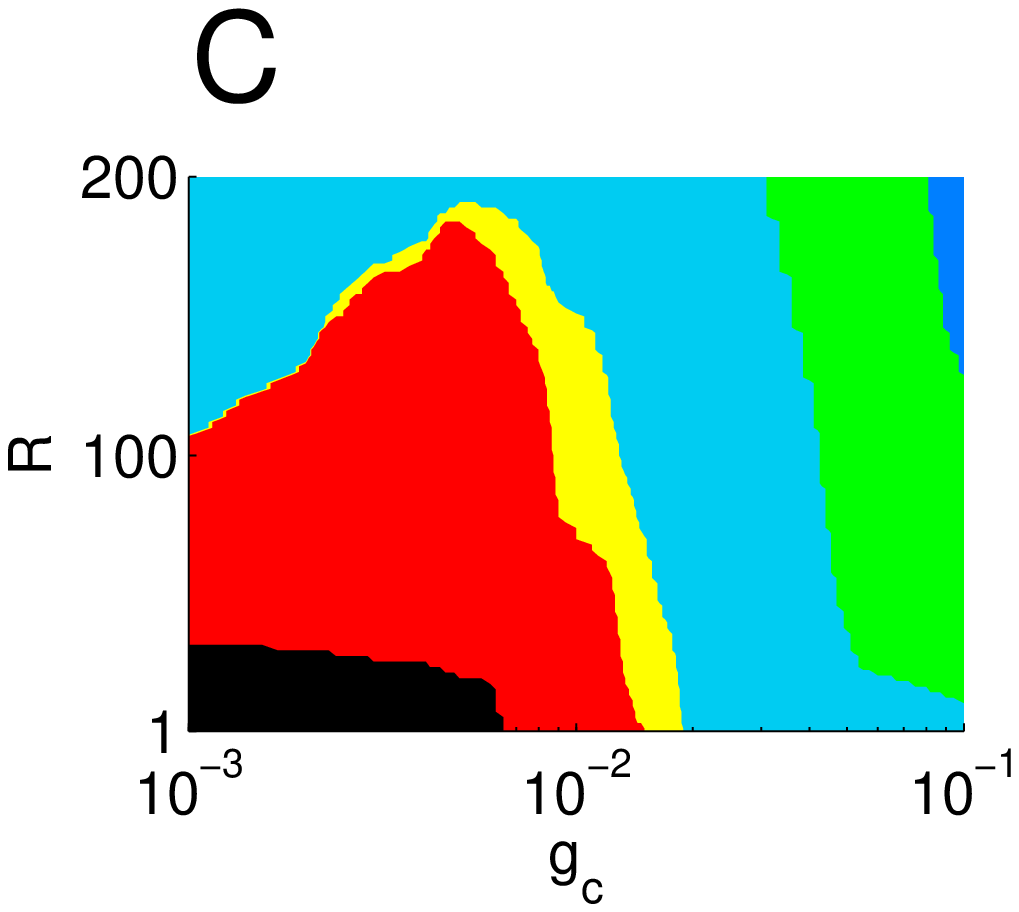}\hspace{00.025cm}
\includegraphics[scale=0.410,trim={10mm 0 5mm 0},clip=true]{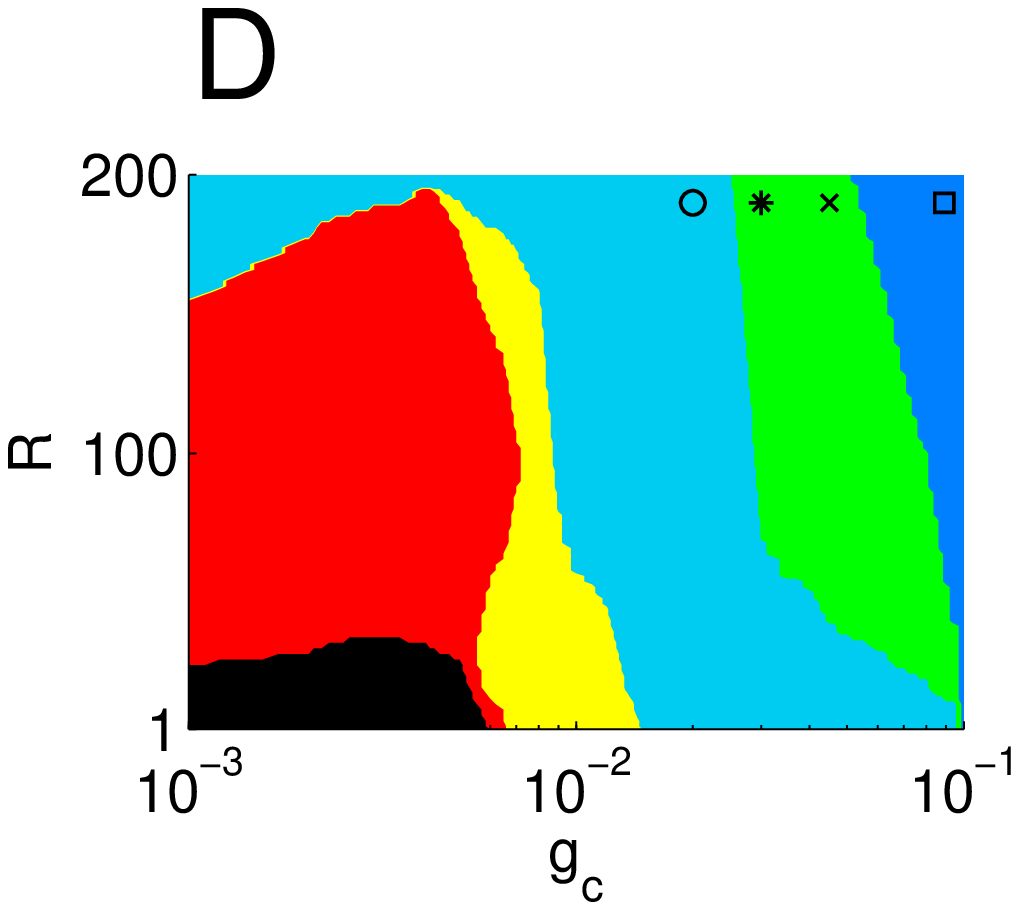}\\
\caption{Joint effect of chemical coupling strength $g_c$ and electrical connection density $R$ on neural population dynamical behavior. We fix electrical coupling strength as $g_e=10^{-6}\,$\textit{mS/cm}$^2$ and use different chemical connection densities set as $S=5$ (A), $S=100$ (B), $S=200$ (C) and $S=300$ (D). In addition to the previous reported dynamical behavior, new emergent dynamical phases as a chaotic amplitude chimera state (green region) and a coherent bursting state (dark blue) are shown. To visualize how system behavior is in these new phases, in panel D (top-right corner) are marked four representative examples, including coherent spiking, chaotic amplitude chimera and coherent bursting states for $R=190,$ which are further illustrated in Fig. \ref{fig6} to visualize the behavioral transition among these new behavioral states. Chemical coupling values for circle, asterisk, cross and square markers are set to $g_c=0.02\,$\textit{mS/cm}$^2$, $g_c=0.03\,$\textit{mS/cm}$^2$, $g_c=0.045\,$\textit{mS/cm}$^2$ and $g_c=0.09\,$\textit{mS/cm}$^2$, respectively.} 
\label{fig5}
\end{figure}

So far, we have extensively explored dynamical states of hybrid coupled population and determined conditions for the emergence of peculiar traveling wave and chimera behaviors on parameter spaces of ($R$, $S$) and ($R$, $g_e$). After carefully inspecting these two-parameter behavioral maps, it can be obviously concluded that ($i$) chemical synapse number $S$ plays the critical role for emergence of the observed behavioral variety  continuously influenced by the presence of the electrical connections in the hybrid coupled population;  ($ii$) this variety can occur only with the presence of relatively weak electrical coupling.

\begin{figure}[t]
\begin{minipage}{0.233\textwidth}
\includegraphics[scale=0.55,trim={8mm 0 40mm 0},clip=true]{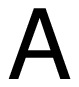}\\
\vspace{0.10cm}
\includegraphics[scale=0.52]{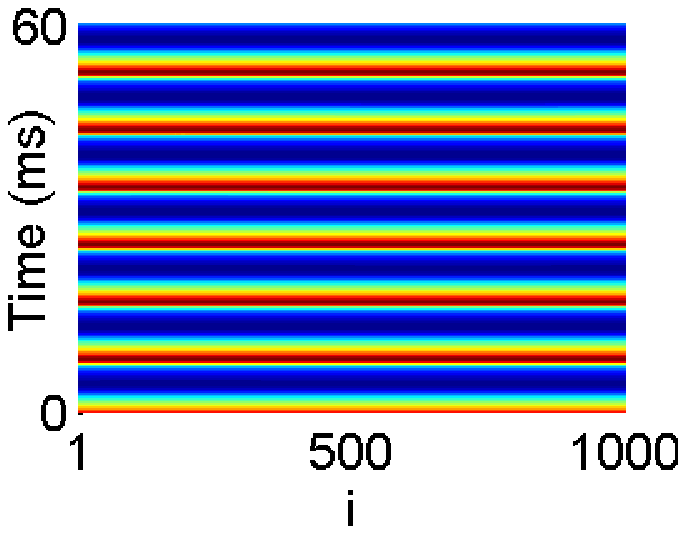}\hspace{-0.00cm}\\
\vspace{0.10cm}
\includegraphics[scale=0.52, angle =0]{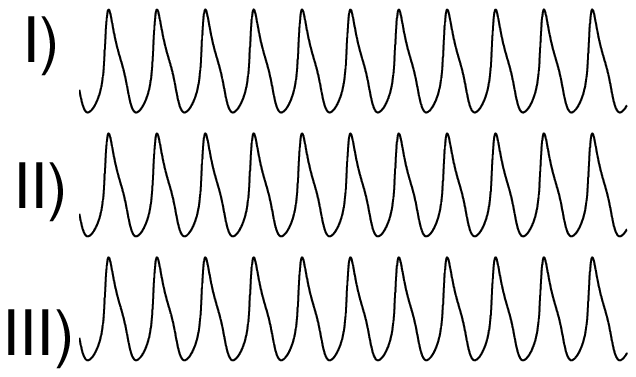}\hspace{-0.00cm}\\
\vspace{-00.00cm}
\end{minipage}
\begin{minipage}{0.233\textwidth}
\includegraphics[scale=0.55,trim={9mm 0 40mm 0},clip=true]{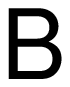}\\
\vspace{0.10cm}
\includegraphics[scale=0.52]{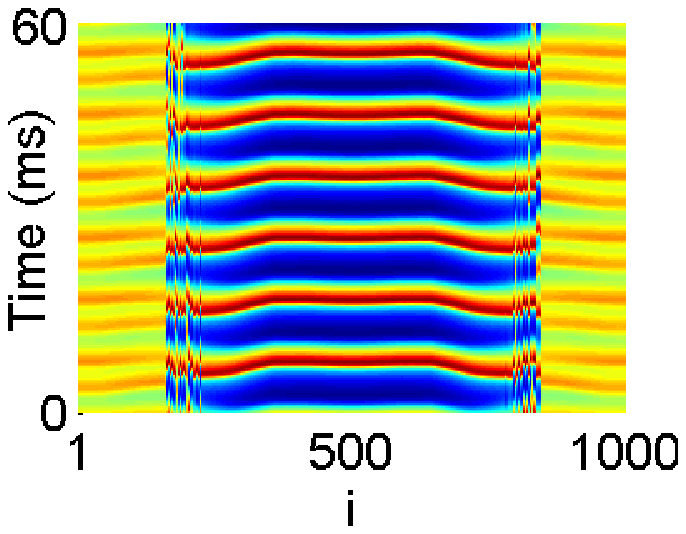}\hspace{-0.00cm}\\
\vspace{0.10cm}
\includegraphics[scale=0.52, angle =0]{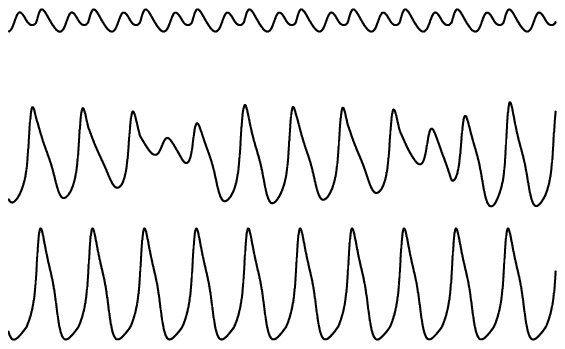}\hspace{-0.00cm}\\
\vspace{-00.00cm}
\end{minipage}
\begin{minipage}{0.233\textwidth}
\includegraphics[scale=0.55,trim={9mm 0 40mm 0},clip=true]{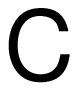}\\
\vspace{0.10cm}
\includegraphics[scale=0.52]{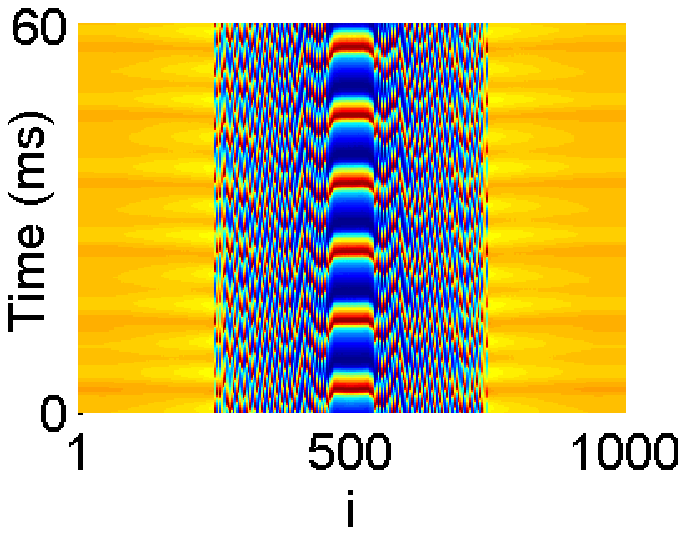}\hspace{-0.00cm}\\
\vspace{0.10cm}
\includegraphics[scale=0.52, angle =0]{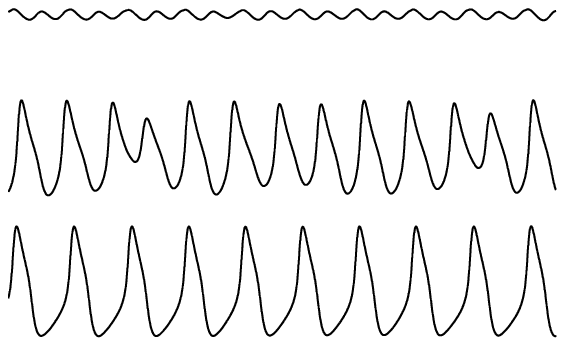}\hspace{-0.00cm}\\
\vspace{-00.00cm}
\end{minipage}
\begin{minipage}{0.25\textwidth}
\includegraphics[scale=0.55,trim={12mm 0 39mm 0},clip=true]{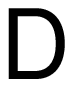}\\
\vspace{0.10cm}
\includegraphics[scale=0.52]{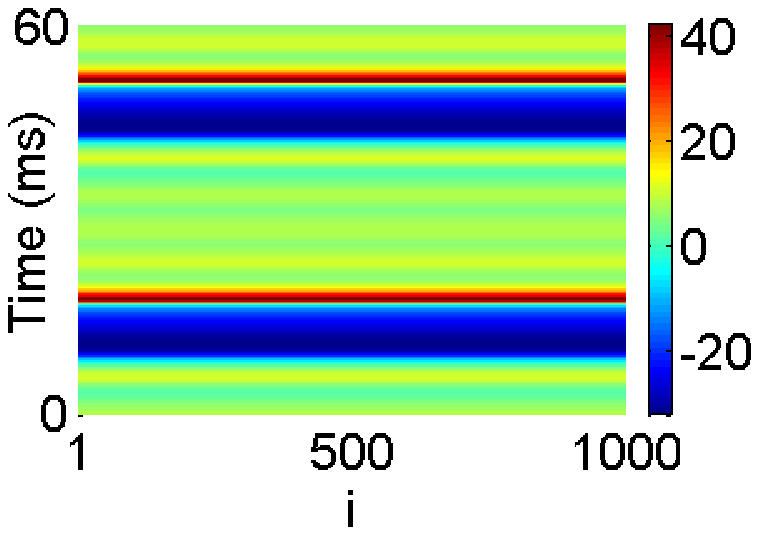}\hspace{-0.0cm}\\
\vspace{0.10cm}
\includegraphics[scale=0.52, angle =0]{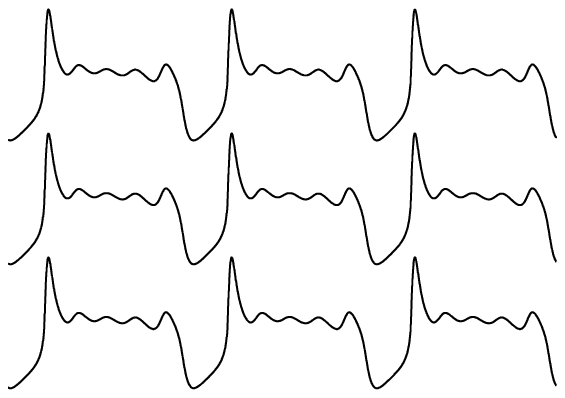}\hspace{-0.0cm}\\
\vspace{-00.00cm}
\end{minipage}
\caption{Representative neural population behaviors are illustrated with respective spatiotemporal patterns (in top panels) corresponding to circle (A), asterisk (B), cross (C) and square (D) markers in Fig.~\ref{fig5}D, respectively. Bottom panels show time series of three neurons randomly selected from different behavioral subgroups, if there are any. System parameters are set as $g_c=0.02\,$\textit{mS/cm}$^2$ (A), $g_c=0.03\,$\textit{mS/cm}$^2$ (B), $g_c=0.045\,$\textit{mS/cm}$^2$ (C) and $g_c=0.09\,$\textit{mS/cm}$^2$ (D). Number of electrical synapses is fixed as $R=190$.}
\label{fig6}
\end{figure}

In order to further understand the effect of the chemical connectivity and to corroborate the robustness of the intriguing emergent dynamical states in the hybrid coupled system, we now continue to investigate how global population behavior might change in the parameter space ($R$, $g_c$) for a moderate electrical coupling strength $g_e=10^{-6}\,$\textit{mS/cm}$^2$ and for distinct values of $S$. Obtained results are depicted in Fig.~\ref{fig5}. Our analysis show that for small $S,$ chemical interactions induce variety of collective dynamical behavior in the neuron population only when $g_c$ gets higher values, depending on the given value of the electrical connection density $R$ (see Fig.~\ref{fig5}A). It is evident that coherent behavior prevails in most regions of ($R$, $g_c$) parameter space and, traveling wave and chimera state only occur at high chemical coupling strengths for small $S$. Increasing $S$ results in enlargement of the regions for traveling wave and chimera state and also shrinkage of the regions for incoherent and coherent states on ($R$, $g_c$) parameter space. On the other hand, we also observe that two new types of complex  behavior start to emerge in population dynamics with the increase in $S$. One is a non-standard chimera state where two different groups of neurons are oscillating coherently with different amplitudes coexisting with an incoherent group of neurons in the same population (see Fig. \ref{fig6}). Here, we call this stable state as \textit{chaotic amplitude chimera} and it occurs at green region in Fig. \ref{fig5}B, C and D. Other behavior is a coherent bursting state (dark blue regions in Fig. \ref{fig5}C and D) where all the neurons in the population exhibit synchronized burst activity instead of regular spiking. These new findings indicate that cooperative effect of electrical and chemical synapses enriches population dynamics inducing new emergent intriguing behaviors, i.e. chaotic amplitude chimera and coherent bursting state, that are not present in populations coupled by only one synapse type.

To better illustrate the dynamical features of the chaotic amplitude chimera and coherent bursting state, we analyze spatiotemporal evolution of the hybrid coupled neural population for four different representative space points marked on ($R$, $g_c$) plane in Fig.~\ref{fig5}D. Our observations are shown in Fig.~\ref{fig6} that also includes the time evaluation of the membrane potentials of randomly selected three neurons from different behavioral subpopulations (bottom panels). It is seen that hybrid coupled population exhibits fully synchronized coherent spiking behavior for the case of $g_c=0.02\,$\textit{mS/cm}$^2$ (see top panel of Fig.~\ref{fig6}A). In this case, the membrane potentials of sample neurons also exhibit a steady-oscillatory spiking behavior as a single cell (bottom panel of Fig.~\ref{fig6}A). Then, an increase in $g_c$ to $0.03\,$\textit{mS/cm}$^2$ (top panel of Fig.~\ref{fig6}B) induces the emergence of the chaotic amplitude chimera in which neurons in the population move on three different basins of attraction, that form two different coherent subpopulations oscillating with different amplitude, and one incoherent subpopulation. This complex population behavior is also confirmed by monitoring the time evolution of the membrane potentials of three cells chosen from each subpopulation (bottom panel of  Fig.~\ref{fig6}B). To check the persistence of this interesting behavior, we further increased $g_c$ and observed that the number of neurons in coherent subpopulation oscillating with large amplitude decreases, while the subpopulation size for the one with small amplitude oscillating and the incoherent group increase (see top panel of Fig.~\ref{fig6}C). Finally, a further increase in $g_c$ to $0.09\,$\textit{mS/cm}$^2$ (Fig.~\ref{fig6}D), there appears a dramatic change in population behavior with the emergence of a fully synchronized bursting type of neural activity with a number of small amplitude oscillations within the burst separated by a single large amplitude oscillation between bursts.

\begin{figure}[b]
\begin{minipage}{0.25\textwidth}
\includegraphics[scale=0.748,trim={3mm 0 20mm 0},clip=true]{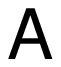}\hspace{-00.00cm}
\includegraphics[scale=0.52]{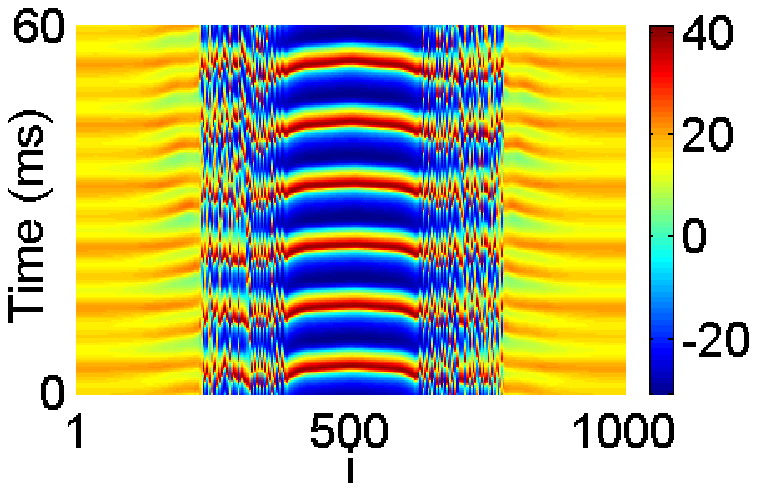}\hspace{-0.00cm}\\
\end{minipage}
\begin{minipage}{0.24\textwidth}
\includegraphics[scale=0.748,trim={4mm 0 20mm 0},clip=true]{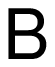}\hspace{-00.00cm}
\includegraphics[scale=0.52]{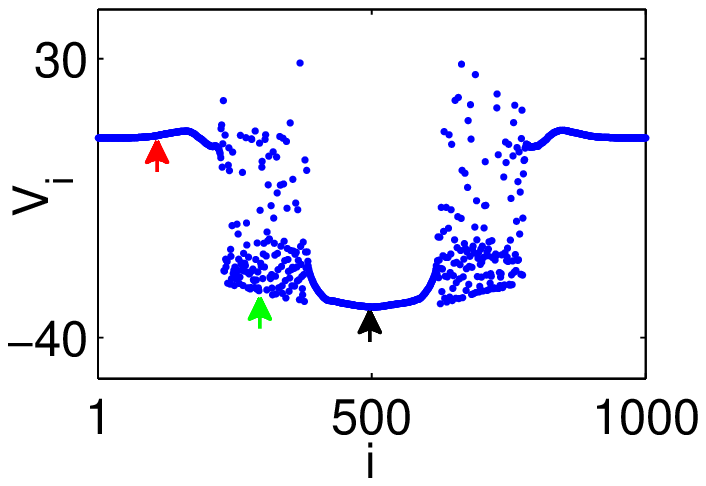}\hspace{-0.00cm}\\
\end{minipage}
\begin{minipage}{0.24\textwidth}
\includegraphics[scale=0.748,trim={4mm 0 20mm 0},clip=true]{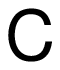}\hspace{-00.00cm}
\includegraphics[scale=0.52]{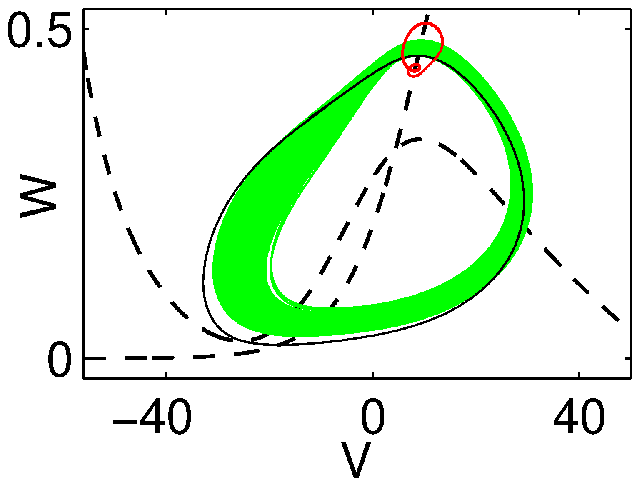}\hspace{-0.00cm}\\
\end{minipage}
\begin{minipage}{0.24\textwidth}
\includegraphics[scale=0.748,trim={4mm 0 20mm 0},clip=true]{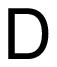}\hspace{-00.00cm}
\includegraphics[scale=0.52]{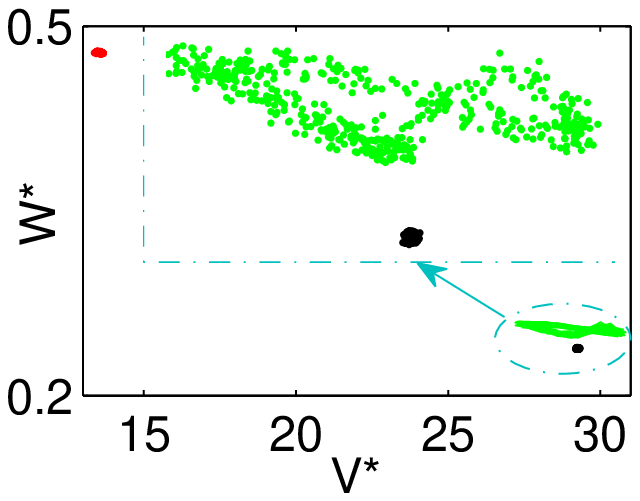}\hspace{-0.00cm}\\
\end{minipage}
\caption{Illustration of a chaotic amplitude chimera. Panels (A) and (B) show spatiotemporal pattern and snapshot of neuron population. Red, green and black arrows in panel (B) show $100^{th}$, $300^{th}$ and $500^{th}$ neurons, respectively. Panels (C) and (D) depict phase plane analysis and Poincare maps (with an enlarged view of the chaotic region) of corresponding neurons. System parameters are set as $S=100$, $R=200$, $g_e=10^{-6}\,$\textit{mS/cm}$^2$ and $g_c=10^{-1}\,$\textit{mS/cm}$^2$.}
\label{fig7}
\end{figure}

Finally, to provide a better understanding and quantitative analysis of the dynamical features of chaotic amplitude chimera state, we illustrate in Fig.~\ref{fig7} different distinguishable views of its behavioral evolution. The figure shows that neural population splits into three domains in such state where different behaviors coexist: small amplitude, large amplitude and incoherent oscillations. We plot spatiotemporal evolution of the hybrid coupled population in Fig.~\ref{fig7}A and give a snapshot of the system in Fig.~\ref{fig7}B. With a careful inspection, one can see that there are actually three different basins of attraction. This can be clearly understood by the phase plane analysis of selected three neurons within different subpopulations. To do this, we randomly choose a representative neuron from each behavioral group which is pointed with red, green and black arrows, and plot corresponding time series on $V$-$w$ plane in Fig.~\ref{fig7}C. It is clearly depicted that there are two different periodic orbits forming distinct coherent behaviors and one unstable periodic orbit resulting in incoherent behavior. In fact, this unstable periodic orbit has a chaotic nature that should be discerned from incoherent state of standard chimeras characterized as in Fig.~\ref{fig2}A where each neuron in the incoherent population follows a non-chaotic stable periodic orbit trajectory. This is also confirmed with the corresponding Poincare maps of the dynamical behavior of the three previous representative neurons illustrated in Fig.~\ref{fig7}D. This clearly shows two distinct coherent behaviors of neurons represented by red and black points plotted in the phase plane corresponding to the low and large amplitude population oscillations and a cloud of points for the chaotic green trajectory. The magnification box makes chaotic behavior of incoherent group more clear.

\section{Discussion}

{Chimera state is a recently discovered dynamical system behavior which has  attracted an increasing interest, and which is characterized by the coexistence of synchronization and desynchronization within a population of identical dynamical elements. This interesting phenomenon has been studied in a wide range of natural and artificial systems, as well as in neuron populations. For the later one, taking into account that the architecture of coupling is a crucial factor for its emergence \cite{omelchenko2015robustness, zhao2018enhancing}, most neural chimera works have focused on networks consisting of either only electrical or only chemical synapses. Although neural chimera emergence has recently been investigated with different hybrid coupling schemes \cite{hizanidis2016chimera, majhi2017chimera, mishra2017traveling}, such studies are very preliminary and make too simplistic assumptions that provide inadequate results when confronted with actual physiological conditions. With the motivation to provide a deeper understanding for the emergence of chimera states in actual neural systems, we here present a comprehensive analysis of how such intriguing dynamical behavior can emerge in a neural system including hybrid synaptic coupling.}

We have first reported the occurrence of chimera-like behaviors in a population of Morris-Lecar neurons, which are coupled by electrical and chemical synapses in a regular network constituting a hybrid coupling scheme. We have explored how dynamical behavior of system changes as a function of the different features of the hybrid connectivity. In particular, we concentrated our analysis on the role of the connection type densities and coupling strengths of electrical and chemical synapses on emergent behavior. It is shown that hybrid coupled populations exhibit variety of dynamical behaviors as a function of electrical and chemical synapse densities in the network. Our findings reveal that chemical synapses, compared to electrical ones, play more significant roles in determining richness of dynamical behavior of the population. Despite this, we also observed that such behavioral variety can only occur in the presence of relatively weak electrical connections. In fact, when electrical coupling strength increases, population exhibits more synchronized behavior as well as the probability to see chimera-like behaviors dramatically decreases. On the other hand, evaluating the effect of chemical coupling strength on population behavior, we found a different trend when the chemical synapse density increases further. In cases of large chemical synaptic strength, the neural population exhibits a new behavior that we have called chaotic amplitude chimera state which has not been reported before. We also observed a pronounced change in membrane potentials of Morris-Lecar neurons for highly intense chemical interaction within hybrid coupled population in such a way that the coherent regular spiking behavior changes to a coherent bursting state. Since pure chemically connected Morris-Lecar neuron population exhibits incoherent, traveling wave, chimera and coherent (spiking) states, we conclude that presence of electrical connections gives rise the emergence of these new intriguing dynamical states.

{Understanding the underlying mechanisms of cognitive processes in actual brains is crucial for appropriate design of the artificially intelligent systems. There are many experimental and theoretical findings that have shown the relation between observed dynamical states in this paper and various cognitive processes \cite{vassileiou2018alignment, mason2014neurocognitive, makeig2002dynamic, klimesch2011alpha, zauner2014lexical}. For instance, synchronization is widely assumed to be a essential mechanism for selective attention \cite{womelsdorf2007role} and memory processes \cite{fell2011role}. Also, it has been shown that traveling waves are closely associated with cognitive processes, ranging from long-term memory consolidation to processing of dynamic visual stimuli \cite{fellinger2012evoked, muller2018cortical}. On the other hand, as chimera state is a recently discovered population behavior, the knowledge of its current role in cognitive processing is still lacking. However, chimera state can naturally appear in brain which satisfies the minimal requirements for its emergence. A well-known example is the unihemispheric sleep activity observed in some marine mammals where their half brain exhibits coherent electrical activity while the other half is incoherent \cite{rattenborg2000behavioral, 10.1371/journal.pone.0217025}. In terms of cognition, chimera state may represent pattern recognition, episodic and spatial memory, similarly to the localized patterns of excitation or ``bump states'' which can also be interpreted as one of dynamical attractors of the working memory \cite{ferreira2016multi, pereira2015tradeoff}. Moreover, one can associate chimeric behavior with event-related synchronization, task switching or multitasking functional states applied in artificially-intelligent systems \cite{adams2018will, valeriani2019brain}.}

{Neural chimera studies may also provide different insights for the understanding of pathological conditions, particularly seizure-related, originating from impairment of balance between synchronous and asynchronous activity. Given the diversity of emergent states reported in our work and the knowledge of the critical conditions for formation and dissolution of chimera states, this knowledge can be useful for an appropriate design of cure strategies of those diseases. For the future studies to investigate chimera state, hybrid coupling concept can be extended to networks including synaptic plasticity with combination of excitatory/inhibitory synapses and gap junctions, and perhaps with different network topologies, i.e. scale-free, small-world and multilayered networks.}

\begin{acknowledgements}
MU acknowledges Bulent Ecevit University Research Foundation under Project No. BAP2018-39971044-01. JJT acknowledges the Spanish Ministry for Science and Technology and the “Agencia Española de Investigación” (AEI) for financial support under grant FIS2017-84256-P (FEDER funds). AC acknowledges financial support from the Scientific and Technological Research Council of Turkey (TUBITAK) BIDEB-2214/A International Research Fellowship Program, and the hospitality of the Institute Carlos I for Theoretical and Computational Physics at University of Granada. 
\end{acknowledgements}

\bibliography{hybridchimera}

\begin{thebibliography}{103}
\expandafter\ifx\csname natexlab\endcsname\relax\def\natexlab#1{#1}\fi
\expandafter\ifx\csname bibnamefont\endcsname\relax
  \def\bibnamefont#1{#1}\fi
\expandafter\ifx\csname bibfnamefont\endcsname\relax
  \def\bibfnamefont#1{#1}\fi
\expandafter\ifx\csname citenamefont\endcsname\relax
  \def\citenamefont#1{#1}\fi
\expandafter\ifx\csname url\endcsname\relax
  \def\url#1{\texttt{#1}}\fi
\expandafter\ifx\csname urlprefix\endcsname\relax\def\urlprefix{URL }\fi
\providecommand{\bibinfo}[2]{#2}
\providecommand{\eprint}[2][]{\url{#2}}

\bibitem[{\citenamefont{Nasir et~al.}(2016)\citenamefont{Nasir, Durrani,
  Mehrpouyan, Blostein, and Kennedy}}]{nasir2016timing}
\bibinfo{author}{\bibfnamefont{A.~A.} \bibnamefont{Nasir}},
  \bibinfo{author}{\bibfnamefont{S.}~\bibnamefont{Durrani}},
  \bibinfo{author}{\bibfnamefont{H.}~\bibnamefont{Mehrpouyan}},
  \bibinfo{author}{\bibfnamefont{S.~D.} \bibnamefont{Blostein}},
  \bibnamefont{and} \bibinfo{author}{\bibfnamefont{R.~A.}
  \bibnamefont{Kennedy}}, \bibinfo{journal}{EURASIP Journal on Wireless
  Communications and Networking} \textbf{\bibinfo{volume}{2016}},
  \bibinfo{pages}{180} (\bibinfo{year}{2016}).

\bibitem[{\citenamefont{Hasan et~al.}(2018)\citenamefont{Hasan, Feng, and
  Tian}}]{hasan2018gnss}
\bibinfo{author}{\bibfnamefont{K.~F.} \bibnamefont{Hasan}},
  \bibinfo{author}{\bibfnamefont{Y.}~\bibnamefont{Feng}}, \bibnamefont{and}
  \bibinfo{author}{\bibfnamefont{Y.-C.} \bibnamefont{Tian}},
  \bibinfo{journal}{IEEE Transactions on Intelligent Transportation Systems}
  \textbf{\bibinfo{volume}{19}}, \bibinfo{pages}{3915} (\bibinfo{year}{2018}).

\bibitem[{\citenamefont{Kanno et~al.}(2017)\citenamefont{Kanno, Hida, Uchida,
  and Bunsen}}]{kanno2017spontaneous}
\bibinfo{author}{\bibfnamefont{K.}~\bibnamefont{Kanno}},
  \bibinfo{author}{\bibfnamefont{T.}~\bibnamefont{Hida}},
  \bibinfo{author}{\bibfnamefont{A.}~\bibnamefont{Uchida}}, \bibnamefont{and}
  \bibinfo{author}{\bibfnamefont{M.}~\bibnamefont{Bunsen}},
  \bibinfo{journal}{Physical Review E} \textbf{\bibinfo{volume}{95}},
  \bibinfo{pages}{052212} (\bibinfo{year}{2017}).

\bibitem[{\citenamefont{Zhang et~al.}(2019)\citenamefont{Zhang, Pan, Yan, Luo,
  Zou, and Xu}}]{zhang2019cluster}
\bibinfo{author}{\bibfnamefont{L.}~\bibnamefont{Zhang}},
  \bibinfo{author}{\bibfnamefont{W.}~\bibnamefont{Pan}},
  \bibinfo{author}{\bibfnamefont{L.}~\bibnamefont{Yan}},
  \bibinfo{author}{\bibfnamefont{B.}~\bibnamefont{Luo}},
  \bibinfo{author}{\bibfnamefont{X.}~\bibnamefont{Zou}}, \bibnamefont{and}
  \bibinfo{author}{\bibfnamefont{M.}~\bibnamefont{Xu}}, \bibinfo{journal}{IEEE
  Journal of Selected Topics in Quantum Electronics}
  \textbf{\bibinfo{volume}{25}}, \bibinfo{pages}{1} (\bibinfo{year}{2019}).

\bibitem[{\citenamefont{Heil et~al.}(2001)\citenamefont{Heil, Fischer,
  Els{\"a}sser, Mulet, and Mirasso}}]{heil2001chaos}
\bibinfo{author}{\bibfnamefont{T.}~\bibnamefont{Heil}},
  \bibinfo{author}{\bibfnamefont{I.}~\bibnamefont{Fischer}},
  \bibinfo{author}{\bibfnamefont{W.}~\bibnamefont{Els{\"a}sser}},
  \bibinfo{author}{\bibfnamefont{J.}~\bibnamefont{Mulet}}, \bibnamefont{and}
  \bibinfo{author}{\bibfnamefont{C.~R.} \bibnamefont{Mirasso}},
  \bibinfo{journal}{Physical Review Letters} \textbf{\bibinfo{volume}{86}},
  \bibinfo{pages}{795} (\bibinfo{year}{2001}).

\bibitem[{\citenamefont{Wallace et~al.}(2000)\citenamefont{Wallace, Yu, Lu, and
  Harrison}}]{wallace2000synchronization}
\bibinfo{author}{\bibfnamefont{I.}~\bibnamefont{Wallace}},
  \bibinfo{author}{\bibfnamefont{D.}~\bibnamefont{Yu}},
  \bibinfo{author}{\bibfnamefont{W.}~\bibnamefont{Lu}}, \bibnamefont{and}
  \bibinfo{author}{\bibfnamefont{R.~G.} \bibnamefont{Harrison}},
  \bibinfo{journal}{Physical Review A} \textbf{\bibinfo{volume}{63}},
  \bibinfo{pages}{013809} (\bibinfo{year}{2000}).

\bibitem[{\citenamefont{Galin et~al.}(2018)\citenamefont{Galin, Borodianskyi,
  Kurin, Shereshevskiy, Vdovicheva, Krasnov, and
  Klushin}}]{galin2018synchronization}
\bibinfo{author}{\bibfnamefont{M.~A.} \bibnamefont{Galin}},
  \bibinfo{author}{\bibfnamefont{E.~A.} \bibnamefont{Borodianskyi}},
  \bibinfo{author}{\bibfnamefont{V.}~\bibnamefont{Kurin}},
  \bibinfo{author}{\bibfnamefont{I.}~\bibnamefont{Shereshevskiy}},
  \bibinfo{author}{\bibfnamefont{N.}~\bibnamefont{Vdovicheva}},
  \bibinfo{author}{\bibfnamefont{V.~M.} \bibnamefont{Krasnov}},
  \bibnamefont{and} \bibinfo{author}{\bibfnamefont{A.}~\bibnamefont{Klushin}},
  \bibinfo{journal}{Physical Review Applied} \textbf{\bibinfo{volume}{9}},
  \bibinfo{pages}{054032} (\bibinfo{year}{2018}).

\bibitem[{\citenamefont{Chitra and Kuriakose}(2008)}]{chitra2008phase}
\bibinfo{author}{\bibfnamefont{R.}~\bibnamefont{Chitra}} \bibnamefont{and}
  \bibinfo{author}{\bibfnamefont{V.}~\bibnamefont{Kuriakose}},
  \bibinfo{journal}{Chaos: An Interdisciplinary Journal of Nonlinear Science}
  \textbf{\bibinfo{volume}{18}}, \bibinfo{pages}{013125}
  (\bibinfo{year}{2008}).

\bibitem[{\citenamefont{Lysak et~al.}(2016)\citenamefont{Lysak, Peskov, Slinko,
  Tyulenin, Bychkov, and Korchak}}]{lysak2016mathematical}
\bibinfo{author}{\bibfnamefont{T.}~\bibnamefont{Lysak}},
  \bibinfo{author}{\bibfnamefont{N.}~\bibnamefont{Peskov}},
  \bibinfo{author}{\bibfnamefont{M.}~\bibnamefont{Slinko}},
  \bibinfo{author}{\bibfnamefont{Y.~P.} \bibnamefont{Tyulenin}},
  \bibinfo{author}{\bibfnamefont{V.~Y.} \bibnamefont{Bychkov}},
  \bibnamefont{and} \bibinfo{author}{\bibfnamefont{V.}~\bibnamefont{Korchak}},
  \bibinfo{journal}{Chemical Engineering Science}
  \textbf{\bibinfo{volume}{144}}, \bibinfo{pages}{7} (\bibinfo{year}{2016}).

\bibitem[{\citenamefont{Nagiev}(2006)}]{nagiev2006coherent}
\bibinfo{author}{\bibfnamefont{T.~M.} \bibnamefont{Nagiev}},
  \emph{\bibinfo{title}{Coherent Synchronized Oxidation Reactions by Hydrogen
  Peroxide}} (\bibinfo{publisher}{Elsevier}, \bibinfo{year}{2006}).

\bibitem[{\citenamefont{Salazar et~al.}(2004)\citenamefont{Salazar, Jansen, and
  Kuzovkov}}]{salazar2004synchronization}
\bibinfo{author}{\bibfnamefont{R.}~\bibnamefont{Salazar}},
  \bibinfo{author}{\bibfnamefont{A.}~\bibnamefont{Jansen}}, \bibnamefont{and}
  \bibinfo{author}{\bibfnamefont{V.}~\bibnamefont{Kuzovkov}},
  \bibinfo{journal}{Physical Review E} \textbf{\bibinfo{volume}{69}},
  \bibinfo{pages}{031604} (\bibinfo{year}{2004}).

\bibitem[{\citenamefont{Nishikawa and Motter}(2015)}]{nishikawa2015comparative}
\bibinfo{author}{\bibfnamefont{T.}~\bibnamefont{Nishikawa}} \bibnamefont{and}
  \bibinfo{author}{\bibfnamefont{A.~E.} \bibnamefont{Motter}},
  \bibinfo{journal}{New Journal of Physics} \textbf{\bibinfo{volume}{17}},
  \bibinfo{pages}{015012} (\bibinfo{year}{2015}).

\bibitem[{\citenamefont{Herzog et~al.}(2017)\citenamefont{Herzog, Hermanstyne,
  Smyllie, and Hastings}}]{herzog2017regulating}
\bibinfo{author}{\bibfnamefont{E.~D.} \bibnamefont{Herzog}},
  \bibinfo{author}{\bibfnamefont{T.}~\bibnamefont{Hermanstyne}},
  \bibinfo{author}{\bibfnamefont{N.~J.} \bibnamefont{Smyllie}},
  \bibnamefont{and} \bibinfo{author}{\bibfnamefont{M.~H.}
  \bibnamefont{Hastings}}, \bibinfo{journal}{Cold Spring Harbor Perspectives in
  Biology} \textbf{\bibinfo{volume}{9}}, \bibinfo{pages}{a027706}
  (\bibinfo{year}{2017}).

\bibitem[{\citenamefont{Buhr and Takahashi}(2013)}]{buhr2013molecular}
\bibinfo{author}{\bibfnamefont{E.~D.} \bibnamefont{Buhr}} \bibnamefont{and}
  \bibinfo{author}{\bibfnamefont{J.~S.} \bibnamefont{Takahashi}}, in
  \emph{\bibinfo{booktitle}{Circadian Clocks}} (\bibinfo{publisher}{Springer},
  \bibinfo{year}{2013}), pp. \bibinfo{pages}{3--27}.

\bibitem[{\citenamefont{Borg et~al.}(2014)\citenamefont{Borg, Ullner, Alagha,
  Alsaedi, Nesbeth, and Zaikin}}]{borg2014complex}
\bibinfo{author}{\bibfnamefont{Y.}~\bibnamefont{Borg}},
  \bibinfo{author}{\bibfnamefont{E.}~\bibnamefont{Ullner}},
  \bibinfo{author}{\bibfnamefont{A.}~\bibnamefont{Alagha}},
  \bibinfo{author}{\bibfnamefont{A.}~\bibnamefont{Alsaedi}},
  \bibinfo{author}{\bibfnamefont{D.}~\bibnamefont{Nesbeth}}, \bibnamefont{and}
  \bibinfo{author}{\bibfnamefont{A.}~\bibnamefont{Zaikin}},
  \bibinfo{journal}{International Journal of Modern Physics B}
  \textbf{\bibinfo{volume}{28}}, \bibinfo{pages}{1430006}
  (\bibinfo{year}{2014}).

\bibitem[{\citenamefont{Wang et~al.}(2010)\citenamefont{Wang, Wang, Liang, Li,
  and Du}}]{wang2010synchronization}
\bibinfo{author}{\bibfnamefont{Y.}~\bibnamefont{Wang}},
  \bibinfo{author}{\bibfnamefont{Z.}~\bibnamefont{Wang}},
  \bibinfo{author}{\bibfnamefont{J.}~\bibnamefont{Liang}},
  \bibinfo{author}{\bibfnamefont{Y.}~\bibnamefont{Li}}, \bibnamefont{and}
  \bibinfo{author}{\bibfnamefont{M.}~\bibnamefont{Du}},
  \bibinfo{journal}{Neurocomputing} \textbf{\bibinfo{volume}{73}},
  \bibinfo{pages}{2532} (\bibinfo{year}{2010}).

\bibitem[{\citenamefont{Li et~al.}(2007)\citenamefont{Li, Chen, and
  Aihara}}]{li2007stochastic}
\bibinfo{author}{\bibfnamefont{C.}~\bibnamefont{Li}},
  \bibinfo{author}{\bibfnamefont{L.}~\bibnamefont{Chen}}, \bibnamefont{and}
  \bibinfo{author}{\bibfnamefont{K.}~\bibnamefont{Aihara}},
  \bibinfo{journal}{BMC Systems Biology} \textbf{\bibinfo{volume}{1}},
  \bibinfo{pages}{6} (\bibinfo{year}{2007}).

\bibitem[{\citenamefont{Sporns et~al.}(2000)\citenamefont{Sporns, Tononi, and
  Edelman}}]{sporns2000connectivity}
\bibinfo{author}{\bibfnamefont{O.}~\bibnamefont{Sporns}},
  \bibinfo{author}{\bibfnamefont{G.}~\bibnamefont{Tononi}}, \bibnamefont{and}
  \bibinfo{author}{\bibfnamefont{G.~M.} \bibnamefont{Edelman}},
  \bibinfo{journal}{Neural Networks} \textbf{\bibinfo{volume}{13}},
  \bibinfo{pages}{909} (\bibinfo{year}{2000}).

\bibitem[{\citenamefont{Womelsdorf and Fries}(2007)}]{womelsdorf2007role}
\bibinfo{author}{\bibfnamefont{T.}~\bibnamefont{Womelsdorf}} \bibnamefont{and}
  \bibinfo{author}{\bibfnamefont{P.}~\bibnamefont{Fries}},
  \bibinfo{journal}{Current Opinion in Neurobiology}
  \textbf{\bibinfo{volume}{17}}, \bibinfo{pages}{154} (\bibinfo{year}{2007}).

\bibitem[{\citenamefont{Velazquez and
  Wennberg}(2009)}]{velazquez2009coordinated}
\bibinfo{author}{\bibfnamefont{J.~L.~P.} \bibnamefont{Velazquez}}
  \bibnamefont{and} \bibinfo{author}{\bibfnamefont{R.}~\bibnamefont{Wennberg}},
  \emph{\bibinfo{title}{Coordinated Activity in the Brain: Measurements and
  Relevance to Brain Function and Behavior}} (\bibinfo{publisher}{Springer
  Science \& Business Media}, \bibinfo{year}{2009}).

\bibitem[{\citenamefont{Kelso et~al.}(2013)\citenamefont{Kelso, Dumas, and
  Tognoli}}]{kelso2013outline}
\bibinfo{author}{\bibfnamefont{J.~S.} \bibnamefont{Kelso}},
  \bibinfo{author}{\bibfnamefont{G.}~\bibnamefont{Dumas}}, \bibnamefont{and}
  \bibinfo{author}{\bibfnamefont{E.}~\bibnamefont{Tognoli}},
  \bibinfo{journal}{Neural Networks} \textbf{\bibinfo{volume}{37}},
  \bibinfo{pages}{120} (\bibinfo{year}{2013}).

\bibitem[{\citenamefont{Khanna and Carmena}(2017)}]{khanna2017beta}
\bibinfo{author}{\bibfnamefont{P.}~\bibnamefont{Khanna}} \bibnamefont{and}
  \bibinfo{author}{\bibfnamefont{J.~M.} \bibnamefont{Carmena}},
  \bibinfo{journal}{eLife} \textbf{\bibinfo{volume}{6}} (\bibinfo{year}{2017}).

\bibitem[{\citenamefont{Oswal et~al.}(2013)\citenamefont{Oswal, Brown, and
  Litvak}}]{oswal2013synchronized}
\bibinfo{author}{\bibfnamefont{A.}~\bibnamefont{Oswal}},
  \bibinfo{author}{\bibfnamefont{P.}~\bibnamefont{Brown}}, \bibnamefont{and}
  \bibinfo{author}{\bibfnamefont{V.}~\bibnamefont{Litvak}},
  \bibinfo{journal}{Current Opinion in Neurology}
  \textbf{\bibinfo{volume}{26}}, \bibinfo{pages}{662} (\bibinfo{year}{2013}).

\bibitem[{\citenamefont{Pollok et~al.}(2012)\citenamefont{Pollok, Krause,
  Martsch, Wach, Schnitzler, and S{\"u}dmeyer}}]{pollok2012motor}
\bibinfo{author}{\bibfnamefont{B.}~\bibnamefont{Pollok}},
  \bibinfo{author}{\bibfnamefont{V.}~\bibnamefont{Krause}},
  \bibinfo{author}{\bibfnamefont{W.}~\bibnamefont{Martsch}},
  \bibinfo{author}{\bibfnamefont{C.}~\bibnamefont{Wach}},
  \bibinfo{author}{\bibfnamefont{A.}~\bibnamefont{Schnitzler}},
  \bibnamefont{and}
  \bibinfo{author}{\bibfnamefont{M.}~\bibnamefont{S{\"u}dmeyer}},
  \bibinfo{journal}{The Journal of Physiology} \textbf{\bibinfo{volume}{590}},
  \bibinfo{pages}{3203} (\bibinfo{year}{2012}).

\bibitem[{\citenamefont{Abd~Hamid et~al.}(2015)\citenamefont{Abd~Hamid, Gall,
  Speck, Antal, and Sabel}}]{abd2015effects}
\bibinfo{author}{\bibfnamefont{A.~I.} \bibnamefont{Abd~Hamid}},
  \bibinfo{author}{\bibfnamefont{C.}~\bibnamefont{Gall}},
  \bibinfo{author}{\bibfnamefont{O.}~\bibnamefont{Speck}},
  \bibinfo{author}{\bibfnamefont{A.}~\bibnamefont{Antal}}, \bibnamefont{and}
  \bibinfo{author}{\bibfnamefont{B.~A.} \bibnamefont{Sabel}},
  \bibinfo{journal}{Frontiers in Neuroscience} \textbf{\bibinfo{volume}{9}},
  \bibinfo{pages}{391} (\bibinfo{year}{2015}).

\bibitem[{\citenamefont{Li et~al.}(2019)\citenamefont{Li, Yao, Xia, Yin, Deng,
  and Yang}}]{li2019adjustment}
\bibinfo{author}{\bibfnamefont{H.}~\bibnamefont{Li}},
  \bibinfo{author}{\bibfnamefont{R.}~\bibnamefont{Yao}},
  \bibinfo{author}{\bibfnamefont{X.}~\bibnamefont{Xia}},
  \bibinfo{author}{\bibfnamefont{G.}~\bibnamefont{Yin}},
  \bibinfo{author}{\bibfnamefont{H.}~\bibnamefont{Deng}}, \bibnamefont{and}
  \bibinfo{author}{\bibfnamefont{P.}~\bibnamefont{Yang}},
  \bibinfo{journal}{Frontiers in Human Neuroscience}
  \textbf{\bibinfo{volume}{13}} (\bibinfo{year}{2019}).

\bibitem[{\citenamefont{Hammond et~al.}(2007)\citenamefont{Hammond, Bergman,
  and Brown}}]{hammond2007pathological}
\bibinfo{author}{\bibfnamefont{C.}~\bibnamefont{Hammond}},
  \bibinfo{author}{\bibfnamefont{H.}~\bibnamefont{Bergman}}, \bibnamefont{and}
  \bibinfo{author}{\bibfnamefont{P.}~\bibnamefont{Brown}},
  \bibinfo{journal}{Trends in Neurosciences} \textbf{\bibinfo{volume}{30}},
  \bibinfo{pages}{357} (\bibinfo{year}{2007}).

\bibitem[{\citenamefont{Uhlhaas and Singer}(2006)}]{uhlhaas2006neural}
\bibinfo{author}{\bibfnamefont{P.~J.} \bibnamefont{Uhlhaas}} \bibnamefont{and}
  \bibinfo{author}{\bibfnamefont{W.}~\bibnamefont{Singer}},
  \bibinfo{journal}{Neuron} \textbf{\bibinfo{volume}{52}}, \bibinfo{pages}{155}
  (\bibinfo{year}{2006}).

\bibitem[{\citenamefont{Batista et~al.}(2010)\citenamefont{Batista, Lopes,
  Viana, and Batista}}]{batista2010delayed}
\bibinfo{author}{\bibfnamefont{C.}~\bibnamefont{Batista}},
  \bibinfo{author}{\bibfnamefont{S.}~\bibnamefont{Lopes}},
  \bibinfo{author}{\bibfnamefont{R.~L.} \bibnamefont{Viana}}, \bibnamefont{and}
  \bibinfo{author}{\bibfnamefont{A.~M.} \bibnamefont{Batista}},
  \bibinfo{journal}{Neural Networks} \textbf{\bibinfo{volume}{23}},
  \bibinfo{pages}{114} (\bibinfo{year}{2010}).

\bibitem[{\citenamefont{Uzuntarla et~al.}(2019)\citenamefont{Uzuntarla, Torres,
  Calim, and Barreto}}]{uzuntarla2019synchronization}
\bibinfo{author}{\bibfnamefont{M.}~\bibnamefont{Uzuntarla}},
  \bibinfo{author}{\bibfnamefont{J.~J.} \bibnamefont{Torres}},
  \bibinfo{author}{\bibfnamefont{A.}~\bibnamefont{Calim}}, \bibnamefont{and}
  \bibinfo{author}{\bibfnamefont{E.}~\bibnamefont{Barreto}},
  \bibinfo{journal}{Neural Networks} \textbf{\bibinfo{volume}{110}},
  \bibinfo{pages}{131} (\bibinfo{year}{2019}).

\bibitem[{\citenamefont{Hanslmayr et~al.}(2012)\citenamefont{Hanslmayr,
  Staudigl, and Fellner}}]{hanslmayr2012oscillatory}
\bibinfo{author}{\bibfnamefont{S.}~\bibnamefont{Hanslmayr}},
  \bibinfo{author}{\bibfnamefont{T.}~\bibnamefont{Staudigl}}, \bibnamefont{and}
  \bibinfo{author}{\bibfnamefont{M.-C.} \bibnamefont{Fellner}},
  \bibinfo{journal}{Frontiers in Human Neuroscience}
  \textbf{\bibinfo{volume}{6}}, \bibinfo{pages}{74} (\bibinfo{year}{2012}).

\bibitem[{\citenamefont{Waschke et~al.}(2019)\citenamefont{Waschke, Tune, and
  Obleser}}]{waschke2019neural}
\bibinfo{author}{\bibfnamefont{L.}~\bibnamefont{Waschke}},
  \bibinfo{author}{\bibfnamefont{S.}~\bibnamefont{Tune}}, \bibnamefont{and}
  \bibinfo{author}{\bibfnamefont{J.}~\bibnamefont{Obleser}},
  \bibinfo{journal}{eLife} \textbf{\bibinfo{volume}{8}},
  \bibinfo{pages}{e51501} (\bibinfo{year}{2019}).

\bibitem[{\citenamefont{Nini et~al.}(1995)\citenamefont{Nini, Feingold, Slovin,
  and Bergman}}]{nini1995neurons}
\bibinfo{author}{\bibfnamefont{A.}~\bibnamefont{Nini}},
  \bibinfo{author}{\bibfnamefont{A.}~\bibnamefont{Feingold}},
  \bibinfo{author}{\bibfnamefont{H.}~\bibnamefont{Slovin}}, \bibnamefont{and}
  \bibinfo{author}{\bibfnamefont{H.}~\bibnamefont{Bergman}},
  \bibinfo{journal}{Journal of Neurophysiology} \textbf{\bibinfo{volume}{74}},
  \bibinfo{pages}{1800} (\bibinfo{year}{1995}).

\bibitem[{\citenamefont{Magill et~al.}(2001)\citenamefont{Magill, Bolam, and
  Bevan}}]{magill2001dopamine}
\bibinfo{author}{\bibfnamefont{P.~J.} \bibnamefont{Magill}},
  \bibinfo{author}{\bibfnamefont{J.~P.} \bibnamefont{Bolam}}, \bibnamefont{and}
  \bibinfo{author}{\bibfnamefont{M.~D.} \bibnamefont{Bevan}},
  \bibinfo{journal}{Neuroscience} \textbf{\bibinfo{volume}{106}},
  \bibinfo{pages}{313} (\bibinfo{year}{2001}).

\bibitem[{\citenamefont{Ostojic}(2014)}]{ostojic2014two}
\bibinfo{author}{\bibfnamefont{S.}~\bibnamefont{Ostojic}},
  \bibinfo{journal}{Nature Neuroscience} \textbf{\bibinfo{volume}{17}},
  \bibinfo{pages}{594} (\bibinfo{year}{2014}).

\bibitem[{\citenamefont{Klimesch et~al.}(1997)\citenamefont{Klimesch,
  Doppelmayr, Pachinger, and Russegger}}]{klimesch1997event}
\bibinfo{author}{\bibfnamefont{W.}~\bibnamefont{Klimesch}},
  \bibinfo{author}{\bibfnamefont{M.}~\bibnamefont{Doppelmayr}},
  \bibinfo{author}{\bibfnamefont{T.}~\bibnamefont{Pachinger}},
  \bibnamefont{and}
  \bibinfo{author}{\bibfnamefont{H.}~\bibnamefont{Russegger}},
  \bibinfo{journal}{Cognitive Brain Research} \textbf{\bibinfo{volume}{6}},
  \bibinfo{pages}{83} (\bibinfo{year}{1997}).

\bibitem[{\citenamefont{Kitajima and Toyota}(2013)}]{kitajima2013decision}
\bibinfo{author}{\bibfnamefont{M.}~\bibnamefont{Kitajima}} \bibnamefont{and}
  \bibinfo{author}{\bibfnamefont{M.}~\bibnamefont{Toyota}},
  \bibinfo{journal}{Biologically Inspired Cognitive Architectures}
  \textbf{\bibinfo{volume}{5}}, \bibinfo{pages}{82} (\bibinfo{year}{2013}).

\bibitem[{\citenamefont{Steriade and McCarley}(2013)}]{steriade2013brainstem}
\bibinfo{author}{\bibfnamefont{M.~M.} \bibnamefont{Steriade}} \bibnamefont{and}
  \bibinfo{author}{\bibfnamefont{R.~W.} \bibnamefont{McCarley}},
  \emph{\bibinfo{title}{Brainstem Control of Wakefulness and Sleep}}
  (\bibinfo{publisher}{Springer Science \& Business Media},
  \bibinfo{year}{2013}).

\bibitem[{\citenamefont{Heinrichs-Graham
  et~al.}(2013)\citenamefont{Heinrichs-Graham, Wilson, Santamaria, Heithoff,
  Torres-Russotto, Hutter-Saunders, Estes, Meza, Mosley, and
  Gendelman}}]{heinrichs2013neuromagnetic}
\bibinfo{author}{\bibfnamefont{E.}~\bibnamefont{Heinrichs-Graham}},
  \bibinfo{author}{\bibfnamefont{T.~W.} \bibnamefont{Wilson}},
  \bibinfo{author}{\bibfnamefont{P.~M.} \bibnamefont{Santamaria}},
  \bibinfo{author}{\bibfnamefont{S.~K.} \bibnamefont{Heithoff}},
  \bibinfo{author}{\bibfnamefont{D.}~\bibnamefont{Torres-Russotto}},
  \bibinfo{author}{\bibfnamefont{J.~A.} \bibnamefont{Hutter-Saunders}},
  \bibinfo{author}{\bibfnamefont{K.~A.} \bibnamefont{Estes}},
  \bibinfo{author}{\bibfnamefont{J.~L.} \bibnamefont{Meza}},
  \bibinfo{author}{\bibfnamefont{R.}~\bibnamefont{Mosley}}, \bibnamefont{and}
  \bibinfo{author}{\bibfnamefont{H.~E.} \bibnamefont{Gendelman}},
  \bibinfo{journal}{Cerebral Cortex} \textbf{\bibinfo{volume}{24}},
  \bibinfo{pages}{2669} (\bibinfo{year}{2013}).

\bibitem[{\citenamefont{Ahn et~al.}(2018)\citenamefont{Ahn, Zauber, Worth,
  Witt, and Rubchinsky}}]{ahn2018neural}
\bibinfo{author}{\bibfnamefont{S.}~\bibnamefont{Ahn}},
  \bibinfo{author}{\bibfnamefont{S.~E.} \bibnamefont{Zauber}},
  \bibinfo{author}{\bibfnamefont{R.~M.} \bibnamefont{Worth}},
  \bibinfo{author}{\bibfnamefont{T.}~\bibnamefont{Witt}}, \bibnamefont{and}
  \bibinfo{author}{\bibfnamefont{L.~L.} \bibnamefont{Rubchinsky}},
  \bibinfo{journal}{Clinical Neurophysiology} \textbf{\bibinfo{volume}{129}},
  \bibinfo{pages}{842} (\bibinfo{year}{2018}).

\bibitem[{\citenamefont{Ahn et~al.}(2013)\citenamefont{Ahn, Rubchinsky, and
  Lapish}}]{ahn2013dynamical}
\bibinfo{author}{\bibfnamefont{S.}~\bibnamefont{Ahn}},
  \bibinfo{author}{\bibfnamefont{L.~L.} \bibnamefont{Rubchinsky}},
  \bibnamefont{and} \bibinfo{author}{\bibfnamefont{C.~C.}
  \bibnamefont{Lapish}}, \bibinfo{journal}{Cerebral Cortex}
  \textbf{\bibinfo{volume}{24}}, \bibinfo{pages}{2553} (\bibinfo{year}{2013}).

\bibitem[{\citenamefont{Laing and Chow}(2001)}]{laing2001stationary}
\bibinfo{author}{\bibfnamefont{C.~R.} \bibnamefont{Laing}} \bibnamefont{and}
  \bibinfo{author}{\bibfnamefont{C.~C.} \bibnamefont{Chow}},
  \bibinfo{journal}{Neural Computation} \textbf{\bibinfo{volume}{13}},
  \bibinfo{pages}{1473} (\bibinfo{year}{2001}).

\bibitem[{\citenamefont{Rattenborg}(2006)}]{rattenborg2006birds}
\bibinfo{author}{\bibfnamefont{N.~C.} \bibnamefont{Rattenborg}},
  \bibinfo{journal}{Naturwissenschaften} \textbf{\bibinfo{volume}{93}},
  \bibinfo{pages}{413} (\bibinfo{year}{2006}).

\bibitem[{\citenamefont{Sakaguchi}(2006)}]{sakaguchi2006instability}
\bibinfo{author}{\bibfnamefont{H.}~\bibnamefont{Sakaguchi}},
  \bibinfo{journal}{Physical Review E} \textbf{\bibinfo{volume}{73}},
  \bibinfo{pages}{031907} (\bibinfo{year}{2006}).

\bibitem[{\citenamefont{Truccolo et~al.}(2014)\citenamefont{Truccolo, Ahmed,
  Harrison, Eskandar, Cosgrove, Madsen, Blum, Potter, Hochberg, and
  Cash}}]{truccolo2014neuronal}
\bibinfo{author}{\bibfnamefont{W.}~\bibnamefont{Truccolo}},
  \bibinfo{author}{\bibfnamefont{O.~J.} \bibnamefont{Ahmed}},
  \bibinfo{author}{\bibfnamefont{M.~T.} \bibnamefont{Harrison}},
  \bibinfo{author}{\bibfnamefont{E.~N.} \bibnamefont{Eskandar}},
  \bibinfo{author}{\bibfnamefont{G.~R.} \bibnamefont{Cosgrove}},
  \bibinfo{author}{\bibfnamefont{J.~R.} \bibnamefont{Madsen}},
  \bibinfo{author}{\bibfnamefont{A.~S.} \bibnamefont{Blum}},
  \bibinfo{author}{\bibfnamefont{N.~S.} \bibnamefont{Potter}},
  \bibinfo{author}{\bibfnamefont{L.~R.} \bibnamefont{Hochberg}},
  \bibnamefont{and} \bibinfo{author}{\bibfnamefont{S.~S.} \bibnamefont{Cash}},
  \bibinfo{journal}{Journal of Neuroscience} \textbf{\bibinfo{volume}{34}},
  \bibinfo{pages}{9927} (\bibinfo{year}{2014}).

\bibitem[{\citenamefont{Fard et~al.}(2015)\citenamefont{Fard, Hollensen,
  Heinke, and Trappenberg}}]{fard2015modeling}
\bibinfo{author}{\bibfnamefont{F.~S.} \bibnamefont{Fard}},
  \bibinfo{author}{\bibfnamefont{P.}~\bibnamefont{Hollensen}},
  \bibinfo{author}{\bibfnamefont{D.}~\bibnamefont{Heinke}}, \bibnamefont{and}
  \bibinfo{author}{\bibfnamefont{T.~P.} \bibnamefont{Trappenberg}},
  \bibinfo{journal}{Neural Networks} \textbf{\bibinfo{volume}{72}},
  \bibinfo{pages}{13} (\bibinfo{year}{2015}).

\bibitem[{\citenamefont{Liou et~al.}(2018)\citenamefont{Liou, Ma, Wenzel, Zhao,
  Baird-Daniel, Smith, Daniel, Emerson, Yuste, Schwartz et~al.}}]{liou2018role}
\bibinfo{author}{\bibfnamefont{J.~Y.} \bibnamefont{Liou}},
  \bibinfo{author}{\bibfnamefont{H.}~\bibnamefont{Ma}},
  \bibinfo{author}{\bibfnamefont{M.}~\bibnamefont{Wenzel}},
  \bibinfo{author}{\bibfnamefont{M.}~\bibnamefont{Zhao}},
  \bibinfo{author}{\bibfnamefont{E.}~\bibnamefont{Baird-Daniel}},
  \bibinfo{author}{\bibfnamefont{E.~H.} \bibnamefont{Smith}},
  \bibinfo{author}{\bibfnamefont{A.}~\bibnamefont{Daniel}},
  \bibinfo{author}{\bibfnamefont{R.}~\bibnamefont{Emerson}},
  \bibinfo{author}{\bibfnamefont{R.}~\bibnamefont{Yuste}},
  \bibinfo{author}{\bibfnamefont{T.~H.} \bibnamefont{Schwartz}},
  \bibnamefont{et~al.}, \bibinfo{journal}{Brain}
  \textbf{\bibinfo{volume}{141}}, \bibinfo{pages}{2083} (\bibinfo{year}{2018}).

\bibitem[{\citenamefont{Kuramoto and
  Battogtokh}(2002)}]{kuramoto2002coexistence}
\bibinfo{author}{\bibfnamefont{Y.}~\bibnamefont{Kuramoto}} \bibnamefont{and}
  \bibinfo{author}{\bibfnamefont{D.}~\bibnamefont{Battogtokh}},
  \bibinfo{journal}{Nonlinear Phenomena in Complex Systems}
  \textbf{\bibinfo{volume}{5}}, \bibinfo{pages}{380} (\bibinfo{year}{2002}).

\bibitem[{\citenamefont{Abrams and Strogatz}(2004)}]{abrams2004chimera}
\bibinfo{author}{\bibfnamefont{D.~M.} \bibnamefont{Abrams}} \bibnamefont{and}
  \bibinfo{author}{\bibfnamefont{S.~H.} \bibnamefont{Strogatz}},
  \bibinfo{journal}{Physical Review Letters} \textbf{\bibinfo{volume}{93}},
  \bibinfo{pages}{174102} (\bibinfo{year}{2004}).

\bibitem[{\citenamefont{Omelchenko et~al.}(2013)\citenamefont{Omelchenko,
  Omelchenko, H\"ovel, and Sch\"oll}}]{omelchenko2013nonlocal}
\bibinfo{author}{\bibfnamefont{I.}~\bibnamefont{Omelchenko}},
  \bibinfo{author}{\bibfnamefont{E.}~\bibnamefont{Omelchenko}},
  \bibinfo{author}{\bibfnamefont{P.}~\bibnamefont{H\"ovel}}, \bibnamefont{and}
  \bibinfo{author}{\bibfnamefont{E.}~\bibnamefont{Sch\"oll}},
  \bibinfo{journal}{Physical Review Letters} \textbf{\bibinfo{volume}{110}},
  \bibinfo{pages}{224101} (\bibinfo{year}{2013}).

\bibitem[{\citenamefont{Omelchenko et~al.}(2015)\citenamefont{Omelchenko,
  Provata, Hizanidis, Sch{\"o}ll, and H{\"o}vel}}]{omelchenko2015robustness}
\bibinfo{author}{\bibfnamefont{I.}~\bibnamefont{Omelchenko}},
  \bibinfo{author}{\bibfnamefont{A.}~\bibnamefont{Provata}},
  \bibinfo{author}{\bibfnamefont{J.}~\bibnamefont{Hizanidis}},
  \bibinfo{author}{\bibfnamefont{E.}~\bibnamefont{Sch{\"o}ll}},
  \bibnamefont{and}
  \bibinfo{author}{\bibfnamefont{P.}~\bibnamefont{H{\"o}vel}},
  \bibinfo{journal}{Physical Review E} \textbf{\bibinfo{volume}{91}},
  \bibinfo{pages}{022917} (\bibinfo{year}{2015}).

\bibitem[{\citenamefont{Bera et~al.}(2016)\citenamefont{Bera, Ghosh, and
  Lakshmanan}}]{bera2016chimera}
\bibinfo{author}{\bibfnamefont{B.~K.} \bibnamefont{Bera}},
  \bibinfo{author}{\bibfnamefont{D.}~\bibnamefont{Ghosh}}, \bibnamefont{and}
  \bibinfo{author}{\bibfnamefont{M.}~\bibnamefont{Lakshmanan}},
  \bibinfo{journal}{Physical Review E} \textbf{\bibinfo{volume}{93}},
  \bibinfo{pages}{012205} (\bibinfo{year}{2016}).

\bibitem[{\citenamefont{Calim et~al.}(2018)\citenamefont{Calim, H{\"o}vel,
  Ozer, and Uzuntarla}}]{calim2018chimera}
\bibinfo{author}{\bibfnamefont{A.}~\bibnamefont{Calim}},
  \bibinfo{author}{\bibfnamefont{P.}~\bibnamefont{H{\"o}vel}},
  \bibinfo{author}{\bibfnamefont{M.}~\bibnamefont{Ozer}}, \bibnamefont{and}
  \bibinfo{author}{\bibfnamefont{M.}~\bibnamefont{Uzuntarla}},
  \bibinfo{journal}{Physical Review E} \textbf{\bibinfo{volume}{98}},
  \bibinfo{pages}{062217} (\bibinfo{year}{2018}).

\bibitem[{\citenamefont{Tsigkri-DeSmedt
  et~al.}(2016)\citenamefont{Tsigkri-DeSmedt, Hizanidis, H{\"o}vel, and
  Provata}}]{tsigkri2016multi}
\bibinfo{author}{\bibfnamefont{N.}~\bibnamefont{Tsigkri-DeSmedt}},
  \bibinfo{author}{\bibfnamefont{J.}~\bibnamefont{Hizanidis}},
  \bibinfo{author}{\bibfnamefont{P.}~\bibnamefont{H{\"o}vel}},
  \bibnamefont{and} \bibinfo{author}{\bibfnamefont{A.}~\bibnamefont{Provata}},
  \bibinfo{journal}{The European Physical Journal Special Topics}
  \textbf{\bibinfo{volume}{225}}, \bibinfo{pages}{1149} (\bibinfo{year}{2016}).

\bibitem[{\citenamefont{Glaze et~al.}(2016)\citenamefont{Glaze, Lewis, and
  Bahar}}]{glaze2016chimera}
\bibinfo{author}{\bibfnamefont{T.~A.} \bibnamefont{Glaze}},
  \bibinfo{author}{\bibfnamefont{S.}~\bibnamefont{Lewis}}, \bibnamefont{and}
  \bibinfo{author}{\bibfnamefont{S.}~\bibnamefont{Bahar}},
  \bibinfo{journal}{Chaos: An Interdisciplinary Journal of Nonlinear Science}
  \textbf{\bibinfo{volume}{26}}, \bibinfo{pages}{083119}
  (\bibinfo{year}{2016}).

\bibitem[{\citenamefont{Bansal et~al.}(2019)\citenamefont{Bansal, Garcia,
  Tompson, Verstynen, Vettel, and Muldoon}}]{bansal2019cognitive}
\bibinfo{author}{\bibfnamefont{K.}~\bibnamefont{Bansal}},
  \bibinfo{author}{\bibfnamefont{J.~O.} \bibnamefont{Garcia}},
  \bibinfo{author}{\bibfnamefont{S.~H.} \bibnamefont{Tompson}},
  \bibinfo{author}{\bibfnamefont{T.}~\bibnamefont{Verstynen}},
  \bibinfo{author}{\bibfnamefont{J.~M.} \bibnamefont{Vettel}},
  \bibnamefont{and} \bibinfo{author}{\bibfnamefont{S.~F.}
  \bibnamefont{Muldoon}}, \bibinfo{journal}{Science Advances}
  \textbf{\bibinfo{volume}{5}}, \bibinfo{pages}{eaau8535}
  (\bibinfo{year}{2019}).

\bibitem[{\citenamefont{Pereda}(2014)}]{pereda2014electrical}
\bibinfo{author}{\bibfnamefont{A.~E.} \bibnamefont{Pereda}},
  \bibinfo{journal}{Nature Reviews Neuroscience} \textbf{\bibinfo{volume}{15}},
  \bibinfo{pages}{250} (\bibinfo{year}{2014}).

\bibitem[{\citenamefont{Connors and Long}(2004)}]{connors2004electrical}
\bibinfo{author}{\bibfnamefont{B.}~\bibnamefont{Connors}} \bibnamefont{and}
  \bibinfo{author}{\bibfnamefont{M.}~\bibnamefont{Long}},
  \bibinfo{journal}{Annual Review of Neuroscience}
  \textbf{\bibinfo{volume}{27}}, \bibinfo{pages}{393} (\bibinfo{year}{2004}).

\bibitem[{\citenamefont{Llinas et~al.}(1974)\citenamefont{Llinas, Baker, and
  Sotelo}}]{llinas1974electrotonic}
\bibinfo{author}{\bibfnamefont{R.}~\bibnamefont{Llinas}},
  \bibinfo{author}{\bibfnamefont{R.}~\bibnamefont{Baker}}, \bibnamefont{and}
  \bibinfo{author}{\bibfnamefont{C.}~\bibnamefont{Sotelo}},
  \bibinfo{journal}{Journal of Neurophysiology} \textbf{\bibinfo{volume}{37}},
  \bibinfo{pages}{560} (\bibinfo{year}{1974}).

\bibitem[{\citenamefont{Christie et~al.}(1989)\citenamefont{Christie, Williams,
  and North}}]{christie1989electrical}
\bibinfo{author}{\bibfnamefont{M.}~\bibnamefont{Christie}},
  \bibinfo{author}{\bibfnamefont{J.}~\bibnamefont{Williams}}, \bibnamefont{and}
  \bibinfo{author}{\bibfnamefont{R.}~\bibnamefont{North}},
  \bibinfo{journal}{Journal of Neuroscience} \textbf{\bibinfo{volume}{9}},
  \bibinfo{pages}{3584} (\bibinfo{year}{1989}).

\bibitem[{\citenamefont{Ma et~al.}(2015)\citenamefont{Ma, Juntti, Hu,
  Huguenard, and Fernald}}]{ma2015electrical}
\bibinfo{author}{\bibfnamefont{Y.}~\bibnamefont{Ma}},
  \bibinfo{author}{\bibfnamefont{S.~A.} \bibnamefont{Juntti}},
  \bibinfo{author}{\bibfnamefont{C.~K.} \bibnamefont{Hu}},
  \bibinfo{author}{\bibfnamefont{J.~R.} \bibnamefont{Huguenard}},
  \bibnamefont{and} \bibinfo{author}{\bibfnamefont{R.~D.}
  \bibnamefont{Fernald}}, \bibinfo{journal}{Proceedings of the National Academy
  of Sciences} \textbf{\bibinfo{volume}{112}}, \bibinfo{pages}{3805}
  (\bibinfo{year}{2015}).

\bibitem[{\citenamefont{Chang et~al.}(1999)\citenamefont{Chang, Gonzalez,
  Pinter, and Balice-Gordon}}]{chang1999gap}
\bibinfo{author}{\bibfnamefont{Q.}~\bibnamefont{Chang}},
  \bibinfo{author}{\bibfnamefont{M.}~\bibnamefont{Gonzalez}},
  \bibinfo{author}{\bibfnamefont{M.~J.} \bibnamefont{Pinter}},
  \bibnamefont{and} \bibinfo{author}{\bibfnamefont{R.~J.}
  \bibnamefont{Balice-Gordon}}, \bibinfo{journal}{Journal of Neuroscience}
  \textbf{\bibinfo{volume}{19}}, \bibinfo{pages}{10813} (\bibinfo{year}{1999}).

\bibitem[{\citenamefont{Li et~al.}(2018)\citenamefont{Li, Mi, Zhang, Wang, and
  Wu}}]{li2018dynamic}
\bibinfo{author}{\bibfnamefont{L.}~\bibnamefont{Li}},
  \bibinfo{author}{\bibfnamefont{Y.}~\bibnamefont{Mi}},
  \bibinfo{author}{\bibfnamefont{W.}~\bibnamefont{Zhang}},
  \bibinfo{author}{\bibfnamefont{D.-H.} \bibnamefont{Wang}}, \bibnamefont{and}
  \bibinfo{author}{\bibfnamefont{S.}~\bibnamefont{Wu}},
  \bibinfo{journal}{Frontiers in Computational Neuroscience}
  \textbf{\bibinfo{volume}{12}}, \bibinfo{pages}{16} (\bibinfo{year}{2018}).

\bibitem[{\citenamefont{Hormuzdi et~al.}(2004)\citenamefont{Hormuzdi, Filippov,
  Mitropoulou, Monyer, and Bruzzone}}]{hormuzdi2004electrical}
\bibinfo{author}{\bibfnamefont{S.~G.} \bibnamefont{Hormuzdi}},
  \bibinfo{author}{\bibfnamefont{M.~A.} \bibnamefont{Filippov}},
  \bibinfo{author}{\bibfnamefont{G.}~\bibnamefont{Mitropoulou}},
  \bibinfo{author}{\bibfnamefont{H.}~\bibnamefont{Monyer}}, \bibnamefont{and}
  \bibinfo{author}{\bibfnamefont{R.}~\bibnamefont{Bruzzone}},
  \bibinfo{journal}{Biochimica et Biophysica Acta (BBA) - Biomembranes}
  \textbf{\bibinfo{volume}{1662}}, \bibinfo{pages}{113} (\bibinfo{year}{2004}).

\bibitem[{\citenamefont{Kennedy}(2016)}]{kennedy2016synaptic}
\bibinfo{author}{\bibfnamefont{M.~B.} \bibnamefont{Kennedy}},
  \bibinfo{journal}{Cold Spring Harbor Perspectives in Biology}
  \textbf{\bibinfo{volume}{8}}, \bibinfo{pages}{a016824}
  (\bibinfo{year}{2016}).

\bibitem[{\citenamefont{Dani et~al.}(2010)\citenamefont{Dani, Huang, Bergan,
  Dulac, and Zhuang}}]{dani2010superresolution}
\bibinfo{author}{\bibfnamefont{A.}~\bibnamefont{Dani}},
  \bibinfo{author}{\bibfnamefont{B.}~\bibnamefont{Huang}},
  \bibinfo{author}{\bibfnamefont{J.}~\bibnamefont{Bergan}},
  \bibinfo{author}{\bibfnamefont{C.}~\bibnamefont{Dulac}}, \bibnamefont{and}
  \bibinfo{author}{\bibfnamefont{X.}~\bibnamefont{Zhuang}},
  \bibinfo{journal}{Neuron} \textbf{\bibinfo{volume}{68}}, \bibinfo{pages}{843}
  (\bibinfo{year}{2010}).

\bibitem[{\citenamefont{Kuo et~al.}(2016)\citenamefont{Kuo, Schwartz, and
  Rieke}}]{kuo2016nonlinear}
\bibinfo{author}{\bibfnamefont{S.~P.} \bibnamefont{Kuo}},
  \bibinfo{author}{\bibfnamefont{G.~W.} \bibnamefont{Schwartz}},
  \bibnamefont{and} \bibinfo{author}{\bibfnamefont{F.}~\bibnamefont{Rieke}},
  \bibinfo{journal}{Neuron} \textbf{\bibinfo{volume}{90}}, \bibinfo{pages}{320}
  (\bibinfo{year}{2016}).

\bibitem[{\citenamefont{Smith and Pereda}(2003)}]{smith2003chemical}
\bibinfo{author}{\bibfnamefont{M.}~\bibnamefont{Smith}} \bibnamefont{and}
  \bibinfo{author}{\bibfnamefont{A.~E.} \bibnamefont{Pereda}},
  \bibinfo{journal}{Proceedings of the National Academy of Sciences}
  \textbf{\bibinfo{volume}{100}}, \bibinfo{pages}{4849} (\bibinfo{year}{2003}).

\bibitem[{\citenamefont{Rash et~al.}(1996)\citenamefont{Rash, Dillman,
  Bilhartz, Duffy, Whalen, and Yasumura}}]{rash1996mixed}
\bibinfo{author}{\bibfnamefont{J.~E.} \bibnamefont{Rash}},
  \bibinfo{author}{\bibfnamefont{R.~K.} \bibnamefont{Dillman}},
  \bibinfo{author}{\bibfnamefont{B.~L.} \bibnamefont{Bilhartz}},
  \bibinfo{author}{\bibfnamefont{H.~S.} \bibnamefont{Duffy}},
  \bibinfo{author}{\bibfnamefont{L.~R.} \bibnamefont{Whalen}},
  \bibnamefont{and} \bibinfo{author}{\bibfnamefont{T.}~\bibnamefont{Yasumura}},
  \bibinfo{journal}{Proceedings of the National Academy of Sciences}
  \textbf{\bibinfo{volume}{93}}, \bibinfo{pages}{4235} (\bibinfo{year}{1996}).

\bibitem[{\citenamefont{Hizanidis et~al.}(2016)\citenamefont{Hizanidis,
  Kouvaris, Zamora-L{\'o}pez, D{\'\i}az-Guilera, and
  Antonopoulos}}]{hizanidis2016chimera}
\bibinfo{author}{\bibfnamefont{J.}~\bibnamefont{Hizanidis}},
  \bibinfo{author}{\bibfnamefont{N.~E.} \bibnamefont{Kouvaris}},
  \bibinfo{author}{\bibfnamefont{G.}~\bibnamefont{Zamora-L{\'o}pez}},
  \bibinfo{author}{\bibfnamefont{A.}~\bibnamefont{D{\'\i}az-Guilera}},
  \bibnamefont{and} \bibinfo{author}{\bibfnamefont{C.~G.}
  \bibnamefont{Antonopoulos}}, \bibinfo{journal}{Scientific Reports}
  \textbf{\bibinfo{volume}{6}}, \bibinfo{pages}{19845} (\bibinfo{year}{2016}).

\bibitem[{\citenamefont{Majhi et~al.}(2017)\citenamefont{Majhi, Perc, and
  Ghosh}}]{majhi2017chimera}
\bibinfo{author}{\bibfnamefont{S.}~\bibnamefont{Majhi}},
  \bibinfo{author}{\bibfnamefont{M.}~\bibnamefont{Perc}}, \bibnamefont{and}
  \bibinfo{author}{\bibfnamefont{D.}~\bibnamefont{Ghosh}},
  \bibinfo{journal}{Chaos: An Interdisciplinary Journal of Nonlinear Science}
  \textbf{\bibinfo{volume}{27}}, \bibinfo{pages}{073109}
  (\bibinfo{year}{2017}).

\bibitem[{\citenamefont{Mishra et~al.}(2017)\citenamefont{Mishra, Saha, Ghosh,
  Osipov, and Dana}}]{mishra2017traveling}
\bibinfo{author}{\bibfnamefont{A.}~\bibnamefont{Mishra}},
  \bibinfo{author}{\bibfnamefont{S.}~\bibnamefont{Saha}},
  \bibinfo{author}{\bibfnamefont{D.}~\bibnamefont{Ghosh}},
  \bibinfo{author}{\bibfnamefont{G.~V.} \bibnamefont{Osipov}},
  \bibnamefont{and} \bibinfo{author}{\bibfnamefont{S.~K.} \bibnamefont{Dana}},
  \bibinfo{journal}{Opera Medica et Physiologica} \textbf{\bibinfo{volume}{3}},
  \bibinfo{pages}{14} (\bibinfo{year}{2017}).

\bibitem[{\citenamefont{Morris and Lecar}(1981)}]{morris1981voltage}
\bibinfo{author}{\bibfnamefont{C.}~\bibnamefont{Morris}} \bibnamefont{and}
  \bibinfo{author}{\bibfnamefont{H.}~\bibnamefont{Lecar}},
  \bibinfo{journal}{Biophysical Journal} \textbf{\bibinfo{volume}{35}},
  \bibinfo{pages}{193} (\bibinfo{year}{1981}).

\bibitem[{\citenamefont{Rinzel and Ermentrout}(1989)}]{Rinzel1989ANE}
\bibinfo{author}{\bibfnamefont{J.}~\bibnamefont{Rinzel}} \bibnamefont{and}
  \bibinfo{author}{\bibfnamefont{G.~B.} \bibnamefont{Ermentrout}}, in
  \emph{\bibinfo{booktitle}{Methods in Neuronal Modeling}}, edited by
  \bibinfo{editor}{\bibfnamefont{C.}~\bibnamefont{Koch}} \bibnamefont{and}
  \bibinfo{editor}{\bibfnamefont{I.}~\bibnamefont{Segev}}
  (\bibinfo{publisher}{MIT Press}, \bibinfo{address}{Cambridge, MA, USA},
  \bibinfo{year}{1989}), pp. \bibinfo{pages}{135--169}.

\bibitem[{\citenamefont{Uzuntarla}(2013)}]{uzuntarla2013inverse}
\bibinfo{author}{\bibfnamefont{M.}~\bibnamefont{Uzuntarla}},
  \bibinfo{journal}{Physics Letters A} \textbf{\bibinfo{volume}{377}},
  \bibinfo{pages}{2585} (\bibinfo{year}{2013}).

\bibitem[{\citenamefont{Shein-Idelson et~al.}(2016)\citenamefont{Shein-Idelson,
  Cohen, Ben-Jacob, and Hanein}}]{shein2016modularity}
\bibinfo{author}{\bibfnamefont{M.}~\bibnamefont{Shein-Idelson}},
  \bibinfo{author}{\bibfnamefont{G.}~\bibnamefont{Cohen}},
  \bibinfo{author}{\bibfnamefont{E.}~\bibnamefont{Ben-Jacob}},
  \bibnamefont{and} \bibinfo{author}{\bibfnamefont{Y.}~\bibnamefont{Hanein}},
  \bibinfo{journal}{PLoS Computational Biology} \textbf{\bibinfo{volume}{12}},
  \bibinfo{pages}{e1004883} (\bibinfo{year}{2016}).

\bibitem[{\citenamefont{Kaiser et~al.}(2007)\citenamefont{Kaiser, Krieger,
  Lodish, and Berk}}]{kaiser2007molecular}
\bibinfo{author}{\bibfnamefont{C.~A.} \bibnamefont{Kaiser}},
  \bibinfo{author}{\bibfnamefont{M.}~\bibnamefont{Krieger}},
  \bibinfo{author}{\bibfnamefont{H.}~\bibnamefont{Lodish}}, \bibnamefont{and}
  \bibinfo{author}{\bibfnamefont{A.}~\bibnamefont{Berk}},
  \emph{\bibinfo{title}{Molecular Cell Biology.}} (\bibinfo{publisher}{WH
  Freeman}, \bibinfo{year}{2007}).

\bibitem[{\citenamefont{O'Donnell et~al.}(2017)\citenamefont{O'Donnell,
  Goncalves, Portera-Cailliau, and Sejnowski}}]{o2017beyond}
\bibinfo{author}{\bibfnamefont{C.}~\bibnamefont{O'Donnell}},
  \bibinfo{author}{\bibfnamefont{J.~T.} \bibnamefont{Goncalves}},
  \bibinfo{author}{\bibfnamefont{C.}~\bibnamefont{Portera-Cailliau}},
  \bibnamefont{and} \bibinfo{author}{\bibfnamefont{T.~J.}
  \bibnamefont{Sejnowski}}, \bibinfo{journal}{eLife}
  \textbf{\bibinfo{volume}{6}}, \bibinfo{pages}{e26724} (\bibinfo{year}{2017}).

\bibitem[{\citenamefont{Avermann et~al.}(2012)\citenamefont{Avermann, Tomm,
  Mateo, Gerstner, and Petersen}}]{avermann2012microcircuits}
\bibinfo{author}{\bibfnamefont{M.}~\bibnamefont{Avermann}},
  \bibinfo{author}{\bibfnamefont{C.}~\bibnamefont{Tomm}},
  \bibinfo{author}{\bibfnamefont{C.}~\bibnamefont{Mateo}},
  \bibinfo{author}{\bibfnamefont{W.}~\bibnamefont{Gerstner}}, \bibnamefont{and}
  \bibinfo{author}{\bibfnamefont{C.~C.} \bibnamefont{Petersen}},
  \bibinfo{journal}{Journal of Neurophysiology} \textbf{\bibinfo{volume}{107}},
  \bibinfo{pages}{3116} (\bibinfo{year}{2012}).

\bibitem[{\citenamefont{Panzeri et~al.}(2001)\citenamefont{Panzeri, Rolls,
  Battaglia, and Lavis}}]{panzeri2001speed}
\bibinfo{author}{\bibfnamefont{S.}~\bibnamefont{Panzeri}},
  \bibinfo{author}{\bibfnamefont{E.~T.} \bibnamefont{Rolls}},
  \bibinfo{author}{\bibfnamefont{F.}~\bibnamefont{Battaglia}},
  \bibnamefont{and} \bibinfo{author}{\bibfnamefont{R.}~\bibnamefont{Lavis}},
  \bibinfo{journal}{Network: Computation in Neural Systems}
  \textbf{\bibinfo{volume}{12}}, \bibinfo{pages}{423} (\bibinfo{year}{2001}).

\bibitem[{\citenamefont{Roth and van Rossum}(2009)}]{inbook}
\bibinfo{author}{\bibfnamefont{A.}~\bibnamefont{Roth}} \bibnamefont{and}
  \bibinfo{author}{\bibfnamefont{M.~C.} \bibnamefont{van Rossum}}, in
  \emph{\bibinfo{booktitle}{Computational Modeling Methods for
  Neuroscientists}} (\bibinfo{publisher}{MIT Press}, \bibinfo{year}{2009}),
  vol.~\bibinfo{volume}{6}, pp. \bibinfo{pages}{139--160}.

\bibitem[{\citenamefont{Zhang et~al.}(2018)\citenamefont{Zhang, Watrous, Patel,
  and Jacobs}}]{zhang2018theta}
\bibinfo{author}{\bibfnamefont{H.}~\bibnamefont{Zhang}},
  \bibinfo{author}{\bibfnamefont{A.~J.} \bibnamefont{Watrous}},
  \bibinfo{author}{\bibfnamefont{A.}~\bibnamefont{Patel}}, \bibnamefont{and}
  \bibinfo{author}{\bibfnamefont{J.}~\bibnamefont{Jacobs}},
  \bibinfo{journal}{Neuron} \textbf{\bibinfo{volume}{98}},
  \bibinfo{pages}{1269} (\bibinfo{year}{2018}).

\bibitem[{\citenamefont{Bertram}(2013)}]{bertram2013neuronal}
\bibinfo{author}{\bibfnamefont{E.~H.} \bibnamefont{Bertram}},
  \bibinfo{journal}{Experimental Neurology} \textbf{\bibinfo{volume}{244}},
  \bibinfo{pages}{67} (\bibinfo{year}{2013}).

\bibitem[{\citenamefont{Kang et~al.}(2019)\citenamefont{Kang, Tian, Huo, and
  Liu}}]{kang2019two}
\bibinfo{author}{\bibfnamefont{L.}~\bibnamefont{Kang}},
  \bibinfo{author}{\bibfnamefont{C.}~\bibnamefont{Tian}},
  \bibinfo{author}{\bibfnamefont{S.}~\bibnamefont{Huo}}, \bibnamefont{and}
  \bibinfo{author}{\bibfnamefont{Z.}~\bibnamefont{Liu}},
  \bibinfo{journal}{Scientific Reports} \textbf{\bibinfo{volume}{9}},
  \bibinfo{pages}{1} (\bibinfo{year}{2019}).

\bibitem[{\citenamefont{Varela et~al.}(2001)\citenamefont{Varela, Lachaux,
  Rodriguez, and Martinerie}}]{varela2001brainweb}
\bibinfo{author}{\bibfnamefont{F.}~\bibnamefont{Varela}},
  \bibinfo{author}{\bibfnamefont{J.-P.} \bibnamefont{Lachaux}},
  \bibinfo{author}{\bibfnamefont{E.}~\bibnamefont{Rodriguez}},
  \bibnamefont{and}
  \bibinfo{author}{\bibfnamefont{J.}~\bibnamefont{Martinerie}},
  \bibinfo{journal}{Nature Reviews Neuroscience} \textbf{\bibinfo{volume}{2}},
  \bibinfo{pages}{229} (\bibinfo{year}{2001}).

\bibitem[{\citenamefont{Abeles et~al.}(1994)\citenamefont{Abeles, Prut,
  Bergman, and Vaadia}}]{abeles1994synchronization}
\bibinfo{author}{\bibfnamefont{M.}~\bibnamefont{Abeles}},
  \bibinfo{author}{\bibfnamefont{Y.}~\bibnamefont{Prut}},
  \bibinfo{author}{\bibfnamefont{H.}~\bibnamefont{Bergman}}, \bibnamefont{and}
  \bibinfo{author}{\bibfnamefont{E.}~\bibnamefont{Vaadia}}, in
  \emph{\bibinfo{booktitle}{Progress in Brain Research}}
  (\bibinfo{publisher}{Elsevier}, \bibinfo{year}{1994}), vol.
  \bibinfo{volume}{102}, pp. \bibinfo{pages}{395--404}.

\bibitem[{\citenamefont{Mamun et~al.}(2015)\citenamefont{Mamun, Mace, Lutman,
  Stein, Liu, Aziz, Vaidyanathan, and Wang}}]{mamun2015movement}
\bibinfo{author}{\bibfnamefont{K.}~\bibnamefont{Mamun}},
  \bibinfo{author}{\bibfnamefont{M.}~\bibnamefont{Mace}},
  \bibinfo{author}{\bibfnamefont{M.}~\bibnamefont{Lutman}},
  \bibinfo{author}{\bibfnamefont{J.}~\bibnamefont{Stein}},
  \bibinfo{author}{\bibfnamefont{X.}~\bibnamefont{Liu}},
  \bibinfo{author}{\bibfnamefont{T.}~\bibnamefont{Aziz}},
  \bibinfo{author}{\bibfnamefont{R.}~\bibnamefont{Vaidyanathan}},
  \bibnamefont{and} \bibinfo{author}{\bibfnamefont{S.}~\bibnamefont{Wang}},
  \bibinfo{journal}{Journal of Neural Engineering}
  \textbf{\bibinfo{volume}{12}}, \bibinfo{pages}{056011}
  (\bibinfo{year}{2015}).

\bibitem[{\citenamefont{Uhlhaas et~al.}(2008)\citenamefont{Uhlhaas, Haenschel,
  Nikoli{\'c}, and Singer}}]{uhlhaas2008role}
\bibinfo{author}{\bibfnamefont{P.~J.} \bibnamefont{Uhlhaas}},
  \bibinfo{author}{\bibfnamefont{C.}~\bibnamefont{Haenschel}},
  \bibinfo{author}{\bibfnamefont{D.}~\bibnamefont{Nikoli{\'c}}},
  \bibnamefont{and} \bibinfo{author}{\bibfnamefont{W.}~\bibnamefont{Singer}},
  \bibinfo{journal}{Schizophrenia Bulletin} \textbf{\bibinfo{volume}{34}},
  \bibinfo{pages}{927} (\bibinfo{year}{2008}).

\bibitem[{\citenamefont{Zhao et~al.}(2018)\citenamefont{Zhao, Sun, and
  Xu}}]{zhao2018enhancing}
\bibinfo{author}{\bibfnamefont{N.}~\bibnamefont{Zhao}},
  \bibinfo{author}{\bibfnamefont{Z.}~\bibnamefont{Sun}}, \bibnamefont{and}
  \bibinfo{author}{\bibfnamefont{W.}~\bibnamefont{Xu}},
  \bibinfo{journal}{Scientific Reports} \textbf{\bibinfo{volume}{8}},
  \bibinfo{pages}{8721} (\bibinfo{year}{2018}).

\bibitem[{\citenamefont{Vassileiou et~al.}(2018)\citenamefont{Vassileiou,
  Meyer, Beese, and Friederici}}]{vassileiou2018alignment}
\bibinfo{author}{\bibfnamefont{B.}~\bibnamefont{Vassileiou}},
  \bibinfo{author}{\bibfnamefont{L.}~\bibnamefont{Meyer}},
  \bibinfo{author}{\bibfnamefont{C.}~\bibnamefont{Beese}}, \bibnamefont{and}
  \bibinfo{author}{\bibfnamefont{A.~D.} \bibnamefont{Friederici}},
  \bibinfo{journal}{Neuroimage} \textbf{\bibinfo{volume}{175}},
  \bibinfo{pages}{286} (\bibinfo{year}{2018}).

\bibitem[{\citenamefont{Mason et~al.}(2014)\citenamefont{Mason, Prat, and
  Just}}]{mason2014neurocognitive}
\bibinfo{author}{\bibfnamefont{R.~A.} \bibnamefont{Mason}},
  \bibinfo{author}{\bibfnamefont{C.~S.} \bibnamefont{Prat}}, \bibnamefont{and}
  \bibinfo{author}{\bibfnamefont{M.~A.} \bibnamefont{Just}},
  \bibinfo{journal}{Cerebral Cortex} \textbf{\bibinfo{volume}{24}},
  \bibinfo{pages}{1474} (\bibinfo{year}{2014}).

\bibitem[{\citenamefont{Makeig et~al.}(2002)\citenamefont{Makeig, Westerfield,
  Jung, Enghoff, Townsend, Courchesne, and Sejnowski}}]{makeig2002dynamic}
\bibinfo{author}{\bibfnamefont{S.}~\bibnamefont{Makeig}},
  \bibinfo{author}{\bibfnamefont{M.}~\bibnamefont{Westerfield}},
  \bibinfo{author}{\bibfnamefont{T.-P.} \bibnamefont{Jung}},
  \bibinfo{author}{\bibfnamefont{S.}~\bibnamefont{Enghoff}},
  \bibinfo{author}{\bibfnamefont{J.}~\bibnamefont{Townsend}},
  \bibinfo{author}{\bibfnamefont{E.}~\bibnamefont{Courchesne}},
  \bibnamefont{and} \bibinfo{author}{\bibfnamefont{T.~J.}
  \bibnamefont{Sejnowski}}, \bibinfo{journal}{Science}
  \textbf{\bibinfo{volume}{295}}, \bibinfo{pages}{690} (\bibinfo{year}{2002}).

\bibitem[{\citenamefont{Klimesch et~al.}(2011)\citenamefont{Klimesch,
  Fellinger, and Freunberger}}]{klimesch2011alpha}
\bibinfo{author}{\bibfnamefont{W.}~\bibnamefont{Klimesch}},
  \bibinfo{author}{\bibfnamefont{R.}~\bibnamefont{Fellinger}},
  \bibnamefont{and}
  \bibinfo{author}{\bibfnamefont{R.}~\bibnamefont{Freunberger}},
  \bibinfo{journal}{Frontiers in Psychology} \textbf{\bibinfo{volume}{2}},
  \bibinfo{pages}{118} (\bibinfo{year}{2011}).

\bibitem[{\citenamefont{Zauner et~al.}(2014)\citenamefont{Zauner, Gruber,
  Himmelsto{\ss}, Lechinger, and Klimesch}}]{zauner2014lexical}
\bibinfo{author}{\bibfnamefont{A.}~\bibnamefont{Zauner}},
  \bibinfo{author}{\bibfnamefont{W.}~\bibnamefont{Gruber}},
  \bibinfo{author}{\bibfnamefont{N.~A.} \bibnamefont{Himmelsto{\ss}}},
  \bibinfo{author}{\bibfnamefont{J.}~\bibnamefont{Lechinger}},
  \bibnamefont{and} \bibinfo{author}{\bibfnamefont{W.}~\bibnamefont{Klimesch}},
  \bibinfo{journal}{Neuroimage} \textbf{\bibinfo{volume}{91}},
  \bibinfo{pages}{252} (\bibinfo{year}{2014}).

\bibitem[{\citenamefont{Fell and Axmacher}(2011)}]{fell2011role}
\bibinfo{author}{\bibfnamefont{J.}~\bibnamefont{Fell}} \bibnamefont{and}
  \bibinfo{author}{\bibfnamefont{N.}~\bibnamefont{Axmacher}},
  \bibinfo{journal}{Nature Reviews Neuroscience} \textbf{\bibinfo{volume}{12}},
  \bibinfo{pages}{105} (\bibinfo{year}{2011}).

\bibitem[{\citenamefont{Fellinger et~al.}(2012)\citenamefont{Fellinger, Gruber,
  Zauner, Freunberger, and Klimesch}}]{fellinger2012evoked}
\bibinfo{author}{\bibfnamefont{R.}~\bibnamefont{Fellinger}},
  \bibinfo{author}{\bibfnamefont{W.}~\bibnamefont{Gruber}},
  \bibinfo{author}{\bibfnamefont{A.}~\bibnamefont{Zauner}},
  \bibinfo{author}{\bibfnamefont{R.}~\bibnamefont{Freunberger}},
  \bibnamefont{and} \bibinfo{author}{\bibfnamefont{W.}~\bibnamefont{Klimesch}},
  \bibinfo{journal}{Neuroimage} \textbf{\bibinfo{volume}{59}},
  \bibinfo{pages}{3379} (\bibinfo{year}{2012}).

\bibitem[{\citenamefont{Muller et~al.}(2018)\citenamefont{Muller, Chavane,
  Reynolds, and Sejnowski}}]{muller2018cortical}
\bibinfo{author}{\bibfnamefont{L.}~\bibnamefont{Muller}},
  \bibinfo{author}{\bibfnamefont{F.}~\bibnamefont{Chavane}},
  \bibinfo{author}{\bibfnamefont{J.}~\bibnamefont{Reynolds}}, \bibnamefont{and}
  \bibinfo{author}{\bibfnamefont{T.~J.} \bibnamefont{Sejnowski}},
  \bibinfo{journal}{Nature Reviews Neuroscience} \textbf{\bibinfo{volume}{19}},
  \bibinfo{pages}{255} (\bibinfo{year}{2018}).

\bibitem[{\citenamefont{Rattenborg et~al.}(2000)\citenamefont{Rattenborg,
  Amlaner, and Lima}}]{rattenborg2000behavioral}
\bibinfo{author}{\bibfnamefont{N.~C.} \bibnamefont{Rattenborg}},
  \bibinfo{author}{\bibfnamefont{C.}~\bibnamefont{Amlaner}}, \bibnamefont{and}
  \bibinfo{author}{\bibfnamefont{S.}~\bibnamefont{Lima}},
  \bibinfo{journal}{Neuroscience \& Biobehavioral Reviews}
  \textbf{\bibinfo{volume}{24}}, \bibinfo{pages}{817} (\bibinfo{year}{2000}).

\bibitem[{\citenamefont{Kendall-Bar et~al.}(2019)\citenamefont{Kendall-Bar,
  Vyssotski, Mukhametov, Siegel, and Lyamin}}]{10.1371/journal.pone.0217025}
\bibinfo{author}{\bibfnamefont{J.~M.} \bibnamefont{Kendall-Bar}},
  \bibinfo{author}{\bibfnamefont{A.~L.} \bibnamefont{Vyssotski}},
  \bibinfo{author}{\bibfnamefont{L.~M.} \bibnamefont{Mukhametov}},
  \bibinfo{author}{\bibfnamefont{J.~M.} \bibnamefont{Siegel}},
  \bibnamefont{and} \bibinfo{author}{\bibfnamefont{O.~I.}
  \bibnamefont{Lyamin}}, \bibinfo{journal}{PLoS ONE}
  \textbf{\bibinfo{volume}{14}}, \bibinfo{pages}{1} (\bibinfo{year}{2019}).

\bibitem[{\citenamefont{Ferreira et~al.}(2016)\citenamefont{Ferreira, Erlhagen,
  and Bicho}}]{ferreira2016multi}
\bibinfo{author}{\bibfnamefont{F.}~\bibnamefont{Ferreira}},
  \bibinfo{author}{\bibfnamefont{W.}~\bibnamefont{Erlhagen}}, \bibnamefont{and}
  \bibinfo{author}{\bibfnamefont{E.}~\bibnamefont{Bicho}},
  \bibinfo{journal}{Physica D: Nonlinear Phenomena}
  \textbf{\bibinfo{volume}{326}}, \bibinfo{pages}{32} (\bibinfo{year}{2016}).

\bibitem[{\citenamefont{Pereira and Wang}(2015)}]{pereira2015tradeoff}
\bibinfo{author}{\bibfnamefont{J.}~\bibnamefont{Pereira}} \bibnamefont{and}
  \bibinfo{author}{\bibfnamefont{X.-J.} \bibnamefont{Wang}},
  \bibinfo{journal}{Cerebral Cortex} \textbf{\bibinfo{volume}{25}},
  \bibinfo{pages}{3586} (\bibinfo{year}{2015}).

\bibitem[{\citenamefont{Adams et~al.}(2018)\citenamefont{Adams,
  Encarna{\c{c}}{\~a}o, Rios-Rinc{\'o}n, and Cook}}]{adams2018will}
\bibinfo{author}{\bibfnamefont{K.}~\bibnamefont{Adams}},
  \bibinfo{author}{\bibfnamefont{P.}~\bibnamefont{Encarna{\c{c}}{\~a}o}},
  \bibinfo{author}{\bibfnamefont{A.~M.} \bibnamefont{Rios-Rinc{\'o}n}},
  \bibnamefont{and} \bibinfo{author}{\bibfnamefont{A.~M.} \bibnamefont{Cook}},
  \bibinfo{journal}{Journal of Human Growth and Development}
  \textbf{\bibinfo{volume}{28}}, \bibinfo{pages}{213} (\bibinfo{year}{2018}).

\bibitem[{\citenamefont{Poli et~al.}(2019)\citenamefont{Poli, Valeriani, and
  Cinel}}]{valeriani2019brain}
\bibinfo{author}{\bibfnamefont{R.}~\bibnamefont{Poli}},
  \bibinfo{author}{\bibfnamefont{D.}~\bibnamefont{Valeriani}},
  \bibnamefont{and} \bibinfo{author}{\bibfnamefont{C.}~\bibnamefont{Cinel}},
  \emph{\bibinfo{title}{Brain-Computer Interfaces for Human Augmentation}}
  (\bibinfo{publisher}{MDPI}, \bibinfo{year}{2019}).

\end{thebibliography}

\end{document}